\ifx\mnmacrosloaded\undefined % MN.TEX (Computer Modern version)
%
% plain TeX single / double column macros for the
% Monthly Notices of Royal Astronomical Society
%
% v1.6  (mn.tex)  --- released 18th September 1995 (A. Woollatt)
% v1.5      "     --- released 25th August 1994 (M. Reed)
% v1.4      "     --- released 22nd February 1994
% v1.3  (mnd.tex) --- released 28th November 1992
% v1.26     "     --- released  1st August 1992
% v1.25     "     --- released 25th February 1992
%
% Copyright Cambridge University Press
%
% > Incorporating special symbol code from laa.sty v1.1 (25th Feb 1991)
%   used with the permission of Springer Verlag.
% > Incorporating parts of mssymb.tex (8th July 1987).
% > Incorporating NewFont.sty v ALPHA patchlevel 8 (16th August 1994).
% > Add footlines, add footnotes in double column (18th September
%   1995).

\catcode `\@=11 % @ signs are letters

\def\@version{1.6}
\def\@verdate{18th September 1995}

% Fonts: Computer Modern / Monotype Times (CUP only)
%
% Font family sizes available:
%   8pt, 9pt, 10pt, 11pt, 14pt and 17pt.
%
% Faces available:
%   \rm, math italic, symbol, \it, \bf, \sl, \tt, \sc, \sf, \cal, \em,
%   \mit and \oldstyle.

% define the typeface in use

\newif\ifprod@font

\ifx\@typeface\undefined
  \def\@typeface{Comp. Modern}\prod@fontfalse
\else
  \prod@fonttrue % We want Times
\fi

\def\newfam{\alloc@8\fam\chardef\sixt@@n} % made not outer

\ifprod@font
\font\fiverm=mtr10 at 5pt
\font\fivebf=mtbx10 at 5pt
\font\fiveit=mtti10 at 5pt
\font\fivesl=mtsl10 at 5pt
\font\fivett=cmtt8 at 5pt     \hyphenchar\fivett=-1
\font\fivecsc=mtcsc10 at 5pt
\font\fivesf=mtss10 at 5pt
\font\fivei=mtmi10 at 5pt      \skewchar\fivei='177
\font\fivesy=mtsy10 at 5pt     \skewchar\fivesy='60

\font\sixrm=mtr10 at 6pt
\font\sixbf=mtbx10 at 6pt
\font\sixit=mtti10 at 6pt
\font\sixsl=mtsl10 at 6pt
\font\sixtt=cmtt8 at 6pt      \hyphenchar\sixtt=-1
\font\sixcsc=mtcsc10 at 6pt
\font\sixsf=mtss10 at 6pt
\font\sixi=mtmi10 at 6pt       \skewchar\sixi='177
\font\sixsy=mtsy10 at 6pt      \skewchar\sixsy='60

\font\sevenrm=mtr10 at 7pt
\font\sevenbf=mtbx10 at 7pt
\font\sevenit=mtti10 at 7pt
\font\sevensl=mtsl10 at 7pt
\font\seventt=cmtt8 at 7pt     \hyphenchar\seventt=-1
\font\sevencsc=mtcsc10 at 7pt
\font\sevensf=mtss10 at 7pt
\font\seveni=mtmi10 at 7pt      \skewchar\seveni='177
\font\sevensy=mtsy10 at 7pt     \skewchar\sevensy='60

\font\eightrm=mtr10 at 8pt
\font\eightbf=mtbx10 at 8pt
\font\eightit=mtti10 at 8pt
\font\eighti=mtmi10 at 8pt      \skewchar\eighti='177
\font\eightsy=mtsy10 at 8pt     \skewchar\eightsy='60
\font\eightsl=mtsl10 at 8pt
\font\eighttt=cmtt8             \hyphenchar\eighttt=-1
\font\eightcsc=mtcsc10 at 8pt
\font\eightsf=mtss10 at 8pt

\font\ninerm=mtr10 at 9pt
\font\ninebf=mtbx10 at 9pt
\font\nineit=mtti10 at 9pt
\font\ninei=mtmi10 at 9pt      \skewchar\ninei='177
\font\ninesy=mtsy10 at 9pt     \skewchar\ninesy='60
\font\ninesl=mtsl10 at 9pt
\font\ninett=cmtt9             \hyphenchar\ninett=-1
\font\ninecsc=mtcsc10 at 9pt
\font\ninesf=mtss10 at 9pt

\font\tenrm=mtr10
\font\tenbf=mtbx10
\font\tenit=mtti10
\font\teni=mtmi10		\skewchar\teni='177
\font\tensy=mtsy10		\skewchar\tensy='60
\font\tenex=cmex10
\font\tensl=mtsl10
\font\tentt=cmtt10		\hyphenchar\tentt=-1
\font\tencsc=mtcsc10
\font\tensf=mtss10

\font\elevenrm=mtr10 at 11pt
\font\elevenbf=mtbx10 at 11pt
\font\elevenit=mtti10 at 11pt
\font\eleveni=mtmi10 at 11pt      \skewchar\eleveni='177
\font\elevensy=mtsy10 at 11pt     \skewchar\elevensy='60
\font\elevensl=mtsl10 at 11pt
\font\eleventt=cmtt10 at 11pt     \hyphenchar\eleventt=-1
\font\elevencsc=mtcsc10 at 11pt
\font\elevensf=mtss10 at 11pt

\font\twelverm=mtr10 at 12pt
\font\twelvebf=mtbx10 at 12pt
\font\twelveit=mtti10 at 12pt
\font\twelvesl=mtsl10 at 12pt
\font\twelvett=cmtt12             \hyphenchar\twelvett=-1
\font\twelvecsc=mtcsc10 at 12pt
\font\twelvesf=mtss10 at 12pt
\font\twelvei=mtmi10 at 12pt      \skewchar\twelvei='177
\font\twelvesy=mtsy10 at 12pt     \skewchar\twelvesy='60

\font\fourteenrm=mtr10 at 14pt
\font\fourteenbf=mtbx10 at 14pt
\font\fourteenit=mtti10 at 14pt
\font\fourteeni=mtmi10 at 14pt      \skewchar\fourteeni='177
\font\fourteensy=mtsy10 at 14pt     \skewchar\fourteensy='60
\font\fourteensl=mtsl10 at 14pt
\font\fourteentt=cmtt12 at 14pt     \hyphenchar\fourteentt=-1
\font\fourteencsc=mtcsc10 at 14pt
\font\fourteensf=mtss10 at 14pt

\font\seventeenrm=mtr10 at 17pt
\font\seventeenbf=mtbx10 at 17pt
\font\seventeenit=mtti10 at 17pt
\font\seventeeni=mtmi10 at 17pt      \skewchar\seventeeni='177
\font\seventeensy=mtsy10 at 17pt     \skewchar\seventeensy='60
\font\seventeensl=mtsl10 at 17pt
\font\seventeentt=cmtt12 at 17pt     \hyphenchar\seventeentt=-1
\font\seventeencsc=mtcsc10 at 17pt
\font\seventeensf=mtss10 at 17pt
\else
\font\fiverm=cmr5
\font\fivei=cmmi5             \skewchar\fivei='177
\font\fivesy=cmsy5            \skewchar\fivesy='60
\font\fivebf=cmbx5

\font\sixrm=cmr6
\font\sixi=cmmi6             \skewchar\sixi='177
\font\sixsy=cmsy6            \skewchar\sixsy='60
\font\sixbf=cmbx6

\font\sevenrm=cmr7
\font\sevenit=cmti7
\font\seveni=cmmi7             \skewchar\seveni='177
\font\sevensy=cmsy7            \skewchar\sevensy='60
\font\sevenbf=cmbx7

\font\eightrm=cmr8
\font\eightbf=cmbx8
\font\eightit=cmti8
\font\eighti=cmmi8			\skewchar\eighti='177
\font\eightsy=cmsy8			\skewchar\eightsy='60
\font\eightsl=cmsl8
\font\eighttt=cmtt8			\hyphenchar\eighttt=-1
\font\eightcsc=cmcsc10 at 8pt
\font\eightsf=cmss8

\font\ninerm=cmr9
\font\ninebf=cmbx9
\font\nineit=cmti9
\font\ninei=cmmi9			\skewchar\ninei='177
\font\ninesy=cmsy9			\skewchar\ninesy='60
\font\ninesl=cmsl9
\font\ninett=cmtt9			\hyphenchar\ninett=-1
\font\ninecsc=cmcsc10 at 9pt
\font\ninesf=cmss9

\font\tenrm=cmr10
\font\tenbf=cmbx10
\font\tenit=cmti10
\font\teni=cmmi10		\skewchar\teni='177
\font\tensy=cmsy10		\skewchar\tensy='60
\font\tenex=cmex10
\font\tensl=cmsl10
\font\tentt=cmtt10		\hyphenchar\tentt=-1
\font\tencsc=cmcsc10
\font\tensf=cmss10

\font\elevenrm=cmr10 scaled \magstephalf
\font\elevenbf=cmbx10 scaled \magstephalf
\font\elevenit=cmti10 scaled \magstephalf
\font\eleveni=cmmi10 scaled \magstephalf	\skewchar\eleveni='177
\font\elevensy=cmsy10 scaled \magstephalf	\skewchar\elevensy='60
\font\elevensl=cmsl10 scaled \magstephalf
\font\eleventt=cmtt10 scaled \magstephalf	\hyphenchar\eleventt=-1
\font\elevencsc=cmcsc10 scaled \magstephalf
\font\elevensf=cmss10 scaled \magstephalf

\font\twelverm=cmr10 scaled \magstep1
\font\twelvebf=cmbx10 scaled \magstep1
\font\twelvei=cmmi10 scaled \magstep1      \skewchar\twelvei='177
\font\twelvesy=cmsy10 scaled \magstep1     \skewchar\twelvesy='60

\font\fourteenrm=cmr10 scaled \magstep2
\font\fourteenbf=cmbx10 scaled \magstep2
\font\fourteenit=cmti10 scaled \magstep2
\font\fourteeni=cmmi10 scaled \magstep2		\skewchar\fourteeni='177
\font\fourteensy=cmsy10 scaled \magstep2	\skewchar\fourteensy='60
\font\fourteensl=cmsl10 scaled \magstep2
\font\fourteentt=cmtt10 scaled \magstep2	\hyphenchar\fourteentt=-1
\font\fourteencsc=cmcsc10 scaled \magstep2
\font\fourteensf=cmss10 scaled \magstep2

\font\seventeenrm=cmr10 scaled \magstep3
\font\seventeenbf=cmbx10 scaled \magstep3
\font\seventeenit=cmti10 scaled \magstep3
\font\seventeeni=cmmi10 scaled \magstep3	\skewchar\seventeeni='177
\font\seventeensy=cmsy10 scaled \magstep3	\skewchar\seventeensy='60
\font\seventeensl=cmsl10 scaled \magstep3
\font\seventeentt=cmtt10 scaled \magstep3	\hyphenchar\seventeentt=-1
\font\seventeencsc=cmcsc10 scaled \magstep3
\font\seventeensf=cmss10 scaled \magstep3
\fi

\def\hexnumber#1{\ifcase#1 0\or1\or2\or3\or4\or5\or6\or7\or8\or9\or
  A\or B\or C\or D\or E\or F\fi}

\def\makestrut{%
  \setbox\strutbox=\hbox{%
    \vrule height.7\baselineskip depth.3\baselineskip width \z@}%
}

\def\baselinestretch{1}
\newskip\tmp@bls

\def\b@ls#1{% set baseline using \baselinestretch as a scale factor
  \tmp@bls=#1\relax
  \baselineskip=#1\relax\makestrut
  \normalbaselineskip=\baselinestretch\tmp@bls
  \normalbaselines
}

\def\nostb@ls#1{% set baseline skip ignoring \baselinestretch
  \normalbaselineskip=#1\relax
  \normalbaselines
  \makestrut
}

% families \itfam, \slfam, \bffam, \ttfam defined in PLAIN.
%
% \itfam is \fam4
% \slfam is \fam5
% \bffam is \fam6
% \ttfam is \fam7

\newfam\scfam  % \fam8
\newfam\sffam  % \fam9

\def\mit{\fam\@ne}
\def\cal{\fam\tw@}
\def\em{\ifdim\fontdimen1\font>\z@ \rm\else\it\fi}

\textfont3=\tenex
\scriptfont3=\tenex
\scriptscriptfont3=\tenex

\setbox0=\hbox{\tenex B} \p@renwd=\wd0 % width of the big left (

\def\eightpoint{% 8^6^5 on 10pt
  \def\rm{\fam0\eightrm}%
  \textfont0=\eightrm \scriptfont0=\sixrm \scriptscriptfont0=\fiverm%
  \textfont1=\eighti  \scriptfont1=\sixi  \scriptscriptfont1=\fivei%
  \textfont2=\eightsy \scriptfont2=\sixsy \scriptscriptfont2=\fivesy%
  \textfont\itfam=\eightit\def\it{\fam\itfam\eightit}%
  \ifprod@font
    \scriptfont\itfam=\sixit
      \scriptscriptfont\itfam=\fiveit
  \else
    \scriptfont\itfam=\eightit
      \scriptscriptfont\itfam=\eightit
  \fi
  \textfont\bffam=\eightbf%
    \scriptfont\bffam=\sixbf%
      \scriptscriptfont\bffam=\fivebf%
  \def\bf{\fam\bffam\eightbf}%
  \textfont\slfam=\eightsl\def\sl{\fam\slfam\eightsl}%
  \ifprod@font
    \scriptfont\slfam=\sixsl
      \scriptscriptfont\slfam=\fivesl
  \else
    \scriptfont\slfam=\eightsl
      \scriptscriptfont\slfam=\eightsl
  \fi
  \textfont\ttfam=\eighttt\def\tt{\fam\ttfam\eighttt}%
  \ifprod@font
    \scriptfont\ttfam=\sixtt
      \scriptscriptfont\ttfam=\fivett
  \else
    \scriptfont\ttfam=\eighttt
      \scriptscriptfont\ttfam=\eighttt
  \fi
  \textfont\scfam=\eightcsc\def\sc{\fam\scfam\eightcsc}%
  \ifprod@font
    \scriptfont\scfam=\sixcsc
      \scriptscriptfont\scfam=\fivecsc
  \else
    \scriptfont\scfam=\eightcsc
      \scriptscriptfont\scfam=\eightcsc
  \fi
  \textfont\sffam=\eightsf\def\sf{\fam\sffam\eightsf}%
  \ifprod@font
    \scriptfont\sffam=\sixsf
      \scriptscriptfont\sffam=\fivesf
  \else
    \scriptfont\sffam=\eightsf
      \scriptscriptfont\sffam=\eightsf
  \fi
  \def\oldstyle{\fam\@ne\eighti}%
  \b@ls{10pt}\rm\@viiipt%
}
\def\@viiipt{}

\def\ninepoint{% 9^6^5 on 11pt (two col) / 12 (single col)
  \def\rm{\fam0\ninerm}%
  \textfont0=\ninerm \scriptfont0=\sixrm \scriptscriptfont0=\fiverm%
  \textfont1=\ninei  \scriptfont1=\sixi  \scriptscriptfont1=\fivei%
  \textfont2=\ninesy \scriptfont2=\sixsy \scriptscriptfont2=\fivesy%
  \textfont\itfam=\nineit\def\it{\fam\itfam\nineit}%
  \ifprod@font
    \scriptfont\itfam=\sixit
      \scriptscriptfont\itfam=\fiveit
  \else
    \scriptfont\itfam=\nineit
      \scriptscriptfont\itfam=\nineit
  \fi
  \textfont\bffam=\ninebf%
    \scriptfont\bffam=\sixbf%
      \scriptscriptfont\bffam=\fivebf%
  \def\bf{\fam\bffam\ninebf}%
  \textfont\slfam=\ninesl\def\sl{\fam\slfam\ninesl}%
  \ifprod@font
    \scriptfont\slfam=\sixsl
      \scriptscriptfont\slfam=\fivesl
  \else
    \scriptfont\slfam=\ninesl
      \scriptscriptfont\slfam=\ninesl
  \fi
  \textfont\ttfam=\ninett\def\tt{\fam\ttfam\ninett}%
  \ifprod@font
    \scriptfont\ttfam=\sixtt
      \scriptscriptfont\ttfam=\fivett
  \else
    \scriptfont\ttfam=\ninett
      \scriptscriptfont\ttfam=\ninett
  \fi
  \textfont\scfam=\ninecsc\def\sc{\fam\scfam\ninecsc}%
  \ifprod@font
    \scriptfont\scfam=\sixcsc
      \scriptscriptfont\scfam=\fivecsc
  \else
    \scriptfont\scfam=\ninecsc
      \scriptscriptfont\scfam=\ninecsc
  \fi
  \textfont\sffam=\ninesf\def\sf{\fam\sffam\ninesf}%
  \ifprod@font
    \scriptfont\sffam=\sixsf
      \scriptscriptfont\sffam=\fivesf
  \else
    \scriptfont\sffam=\ninesf
      \scriptscriptfont\sffam=\ninesf
  \fi
  \def\oldstyle{\fam\@ne\ninei}%
  \b@ls{\TextLeading plus \Feathering}\rm\@ixpt%
}
\def\@ixpt{}

\def\tenpoint{% 10^7^5 on 11pt
  \def\rm{\fam0\tenrm}%
  \textfont0=\tenrm \scriptfont0=\sevenrm \scriptscriptfont0=\fiverm%
  \textfont1=\teni  \scriptfont1=\seveni  \scriptscriptfont1=\fivei%
  \textfont2=\tensy \scriptfont2=\sevensy \scriptscriptfont2=\fivesy%
  \textfont\itfam=\tenit\def\it{\fam\itfam\tenit}%
  \ifprod@font
    \scriptfont\itfam=\sevenit
      \scriptscriptfont\itfam=\fiveit
  \else
    \scriptfont\itfam=\tenit
      \scriptscriptfont\itfam=\tenit
  \fi
  \textfont\bffam=\tenbf%
    \scriptfont\bffam=\sevenbf%
      \scriptscriptfont\bffam=\fivebf%
  \def\bf{\fam\bffam\tenbf}%
  \textfont\slfam=\tensl\def\sl{\fam\slfam\tensl}%
  \ifprod@font
    \scriptfont\slfam=\sevensl
      \scriptscriptfont\slfam=\fivesl
  \else
    \scriptfont\slfam=\tensl
      \scriptscriptfont\slfam=\tensl
  \fi
  \textfont\ttfam=\tentt\def\tt{\fam\ttfam\tentt}%
  \ifprod@font
    \scriptfont\ttfam=\seventt
      \scriptscriptfont\ttfam=\fivett
  \else
    \scriptfont\ttfam=\tentt
      \scriptscriptfont\ttfam=\tentt
  \fi
  \textfont\scfam=\tencsc\def\sc{\fam\scfam\tencsc}%
  \ifprod@font
    \scriptfont\scfam=\sevencsc
      \scriptscriptfont\scfam=\fivecsc
  \else
    \scriptfont\scfam=\tencsc
      \scriptscriptfont\scfam=\tencsc
  \fi
  \textfont\sffam=\tensf\def\sf{\fam\sffam\tensf}%
  \ifprod@font
    \scriptfont\sffam=\sevensf
      \scriptscriptfont\sffam=\fivesf
  \else
    \scriptfont\sffam=\tensf
      \scriptscriptfont\sffam=\tensf
  \fi
  \def\oldstyle{\fam\@ne\teni}%
  \b@ls{11pt}\rm\@xpt%
}
\def\@xpt{}

\def\elevenpoint{% 11^8^6 on 13pt
  \def\rm{\fam0\elevenrm}%
  \textfont0=\elevenrm \scriptfont0=\eightrm \scriptscriptfont0=\sixrm%
  \textfont1=\eleveni  \scriptfont1=\eighti  \scriptscriptfont1=\sixi%
  \textfont2=\elevensy \scriptfont2=\eightsy \scriptscriptfont2=\sixsy%
  \textfont\itfam=\elevenit\def\it{\fam\itfam\elevenit}%
  \ifprod@font
    \scriptfont\itfam=\eightit
      \scriptscriptfont\itfam=\sixit
  \else
    \scriptfont\itfam=\elevenit
      \scriptscriptfont\itfam=\elevenit
  \fi
  \textfont\bffam=\elevenbf%
    \scriptfont\bffam=\eightbf%
      \scriptscriptfont\bffam=\sixbf%
  \def\bf{\fam\bffam\elevenbf}%
  \textfont\slfam=\elevensl\def\sl{\fam\slfam\elevensl}%
  \ifprod@font
    \scriptfont\slfam=\eightsl
      \scriptscriptfont\slfam=\sixsl
  \else
    \scriptfont\slfam=\elevensl
      \scriptscriptfont\slfam=\elevensl
  \fi
  \textfont\ttfam=\eleventt\def\tt{\fam\ttfam\eleventt}%
  \ifprod@font
    \scriptfont\ttfam=\eighttt
      \scriptscriptfont\ttfam=\sixtt
  \else
    \scriptfont\ttfam=\eleventt
      \scriptscriptfont\ttfam=\eleventt
  \fi
  \textfont\scfam=\elevencsc\def\sc{\fam\scfam\elevencsc}%
  \ifprod@font
    \scriptfont\scfam=\eightcsc
      \scriptscriptfont\scfam=\sixcsc
  \else
    \scriptfont\scfam=\elevencsc
      \scriptscriptfont\scfam=\elevencsc
  \fi
  \textfont\sffam=\elevensf\def\sf{\fam\sffam\elevensf}%
  \ifprod@font
    \scriptfont\sffam=\eightsf
      \scriptscriptfont\sffam=\sixsf
  \else
    \scriptfont\sffam=\elevensf
      \scriptscriptfont\sffam=\elevensf
  \fi
  \def\oldstyle{\fam\@ne\eleveni}%
  \b@ls{13pt}\rm\@xipt%
}
\def\@xipt{}

\def\fourteenpoint{% 14^10^7 on 17pt
  \def\rm{\fam0\fourteenrm}%
  \textfont0\fourteenrm  \scriptfont0\tenrm  \scriptscriptfont0\sevenrm%
  \textfont1\fourteeni   \scriptfont1\teni   \scriptscriptfont1\seveni%
  \textfont2\fourteensy  \scriptfont2\tensy  \scriptscriptfont2\sevensy%
  \textfont\itfam=\fourteenit\def\it{\fam\itfam\fourteenit}%
  \ifprod@font
    \scriptfont\itfam=\tenit
      \scriptscriptfont\itfam=\sevenit
  \else
    \scriptfont\itfam=\fourteenit
      \scriptscriptfont\itfam=\fourteenit
  \fi
  \textfont\bffam=\fourteenbf%
    \scriptfont\bffam=\tenbf%
      \scriptscriptfont\bffam=\sevenbf%
  \def\bf{\fam\bffam\fourteenbf}%
  \textfont\slfam=\fourteensl\def\sl{\fam\slfam\fourteensl}%
  \ifprod@font
    \scriptfont\slfam=\tensl
      \scriptscriptfont\slfam=\sevensl
  \else
    \scriptfont\slfam=\fourteensl
      \scriptscriptfont\slfam=\fourteensl
  \fi
  \textfont\ttfam=\fourteentt\def\tt{\fam\ttfam\fourteentt}%
  \ifprod@font
    \scriptfont\ttfam=\tentt
      \scriptscriptfont\ttfam=\seventt
  \else
    \scriptfont\ttfam=\fourteentt
      \scriptscriptfont\ttfam=\fourteentt
  \fi
  \textfont\scfam=\fourteencsc\def\sc{\fam\scfam\fourteencsc}%
  \ifprod@font
    \scriptfont\scfam=\tencsc
      \scriptscriptfont\scfam=\sevencsc
  \else
    \scriptfont\scfam=\fourteencsc
      \scriptscriptfont\scfam=\fourteencsc
  \fi
  \textfont\sffam=\fourteensf\def\sf{\fam\sffam\fourteensf}%
  \ifprod@font
    \scriptfont\sffam=\tensf
      \scriptscriptfont\sffam=\sevensf
  \else
    \scriptfont\sffam=\fourteensf
      \scriptscriptfont\sffam=\fourteensf
  \fi
  \def\oldstyle{\fam\@ne\fourteeni}%
  \b@ls{17pt}\rm\@xivpt%
}
\def\@xivpt{}

\def\seventeenpoint{% 17^12^10 on 20pt
  \def\rm{\fam0\seventeenrm}%
  \textfont0\seventeenrm  \scriptfont0\twelverm  \scriptscriptfont0\tenrm%
  \textfont1\seventeeni   \scriptfont1\twelvei   \scriptscriptfont1\teni%
  \textfont2\seventeensy  \scriptfont2\twelvesy  \scriptscriptfont2\tensy%
  \textfont\itfam=\seventeenit\def\it{\fam\itfam\seventeenit}%
  \ifprod@font
    \scriptfont\itfam=\twelveit
      \scriptscriptfont\itfam=\tenit
  \else
    \scriptfont\itfam=\seventeenit
      \scriptscriptfont\itfam=\seventeenit
  \fi
  \textfont\bffam=\seventeenbf%
    \scriptfont\bffam=\twelvebf%
      \scriptscriptfont\bffam=\tenbf%
  \def\bf{\fam\bffam\seventeenbf}%
  \textfont\slfam=\seventeensl\def\sl{\fam\slfam\seventeensl}%
  \ifprod@font
    \scriptfont\slfam=\twelvesl
      \scriptscriptfont\slfam=\tensl
  \else
    \scriptfont\slfam=\seventeensl
      \scriptscriptfont\slfam=\seventeensl
  \fi
  \textfont\ttfam=\seventeentt\def\tt{\fam\ttfam\seventeentt}%
  \ifprod@font
    \scriptfont\ttfam=\twelvett
      \scriptscriptfont\ttfam=\tentt
  \else
    \scriptfont\ttfam=\seventeentt
      \scriptscriptfont\ttfam=\seventeentt
  \fi
  \textfont\scfam=\seventeencsc\def\sc{\fam\scfam\seventeencsc}%
  \ifprod@font
    \scriptfont\scfam=\twelvecsc
      \scriptscriptfont\scfam=\tencsc
  \else
    \scriptfont\scfam=\seventeencsc
      \scriptscriptfont\scfam=\seventeencsc
  \fi
  \textfont\sffam=\seventeensf\def\sf{\fam\sffam\seventeensf}%
  \ifprod@font
    \scriptfont\sffam=\twelvesf
      \scriptscriptfont\sffam=\tensf
  \else
    \scriptfont\sffam=\seventeensf
      \scriptscriptfont\sffam=\seventeensf
  \fi
  \def\oldstyle{\fam\@ne\seventeeni}%
  \b@ls{20pt}\rm\@xviipt%
}
\def\@xviipt{}

\lineskip=1pt      \normallineskip=\lineskip
\lineskiplimit=\z@ \normallineskiplimit=\lineskiplimit

% BOLD MATH SYMBOLS

% Make \, work in non-math mode
\def\,{\relax\ifmmode \mskip\thinmuskip\else \thinspace\fi}
\let\protect=\relax

\long\def\@ifundefined#1#2#3{\expandafter\ifx\csname
  #1\endcsname\relax#2\else#3\fi}

%%%%%%%%%%%%%%%%%%%%%%%%%%%%%%%%%%%%%%%%%

% NewFont.sty: ALPHA VERSION patchlevel 8, 16th August 1994, M. Reed

% \addtom@thgroup{math font loading info}
% Adds to internal \math@groups definition, which is executed at the end
% of each size changing command. It is called by \NewSymbolFont.

\newtoks\math@groups \math@groups={}
\def\addtom@thgroup#1#2{#1\expandafter{\the#1#2}} %  \mac={new\the\mac}

% Make TeX change the values of \s@ze, \ss@ze, \sss@ze when \@npt is
% executed. This makes it possible for math characters to be loaded
% at the correct size automatically when the size is changed.

% \addtosizeh@ok{x}{10}{7}{5}

\def\addtosizeh@ok#1#2#3#4{%
  \expandafter\def\csname @#1pt\endcsname{%
    \def\s@ze{#2}\def\ss@ze{#3}\def\sss@ze{#4}\the\math@groups%
  }%
}

% \resetsizehook allows the size parameters to be reset after \addtosizeh@ok
% has been called (it re-defines \@npt).
% e.g JFM which requires \xpt to have 10.5pt instead of 10pt.
% Note: \resetsizehook must be used in the preamble BEFORE any
% \New... commands.

% e.g. \resetsizehook{x}{10.5}{7}{5}

\let\resetsizehook=\addtosizeh@ok

% Standard LaTeX sizes

\ifprod@font
%  \addtosizeh@ok{v}    {5} {5}  {5}
%  \addtosizeh@ok{vi}   {6} {6}  {6}
%  \addtosizeh@ok{vii}  {7} {6}  {5}
  \addtosizeh@ok{viii} {8} {6}  {5}
  \addtosizeh@ok{ix}   {9} {6}  {5}
  \addtosizeh@ok{x}    {10}{7}  {5}
  \addtosizeh@ok{xi}   {11}{8}  {6}
%  \addtosizeh@ok{xii}  {12}{8}  {6}
  \addtosizeh@ok{xiv}  {14}{10} {7}
  \addtosizeh@ok{xvii} {17}{12}{10}
%  \addtosizeh@ok{xx}   {20}{14}{12}
%  \addtosizeh@ok{xxv}  {25}{20}{17}
\else
%  \addtosizeh@ok{v}    {5}     {5}     {5}
%  \addtosizeh@ok{vi}   {6}     {6}     {6}
%  \addtosizeh@ok{vii}  {7}     {6}     {5}
  \addtosizeh@ok{viii} {8}     {6}     {5}
  \addtosizeh@ok{ix}   {9}     {6}     {5}
  \addtosizeh@ok{x}    {10}    {7}     {5}
  \addtosizeh@ok{xi}   {10.95} {8}     {6}
%  \addtosizeh@ok{xii}  {12}    {8}     {6}
  \addtosizeh@ok{xiv}  {14.4}  {10}    {7}
  \addtosizeh@ok{xvii} {17.28} {12}    {10}
%  \addtosizeh@ok{xx}   {20.74} {14.4}  {12}
%  \addtosizeh@ok{xxv}  {24.88} {20.74} {17.28}
\fi

\def\get@font#1#2#3{%
  \edef\fonts@ze{\romannumeral#3}%         10 -> x
  \edef\fontn@me{\fonts@ze#1}%             AMSa -> xAMSa
  \@ifundefined{\fontn@me}%
    {%%\typeout{defining \fontn@me}%
     \global\expandafter\font\csname \fontn@me\endcsname=#2 at #3pt}%
    {}%
}

\def\ass@tfont#1#2{%
  \xdef\fam@name{\csname #1\endcsname}%
  \xdef\font@name{\csname #2\endcsname}%
  \let\textfont@name\font@name
  \textfont\fam@name\textfont@name
}

\def\ass@sfont#1#2{%
  \xdef\fam@name{\csname #1\endcsname}%
  \xdef\font@name{\csname #2\endcsname}%
  \let\textfont@name\font@name
  \scriptfont\fam@name\textfont@name
}

\def\ass@ssfont#1#2{%
  \xdef\fam@name{\csname #1\endcsname}%
  \xdef\font@name{\csname #2\endcsname}%
  \let\textfont@name\font@name
  \scriptscriptfont\fam@name\textfont@name
}

%                fam name  base font  (allocates a \newfam)
% \NewSymbolFont {AMSa}    {mtxm10}

\def\NewSymbolFont#1#2{%
  \expandafter\ifx\csname sym#1fam\endcsname\relax % if not defined
    \expandafter\newfam\csname sym#1fam\endcsname
    \expandafter\edef\csname sym#1fam\endcsname{\the\allocationnumber}%
    \addtom@thgroup\math@groups{%
      \get@font{#1}{#2}{\s@ze}%
      \ass@tfont{sym#1fam}{\fontn@me}%
      \get@font{#1}{#2}{\ss@ze}%
      \ass@sfont{sym#1fam}{\fontn@me}%
      \get@font{#1}{#2}{\sss@ze}%
      \ass@ssfont{sym#1fam}{\fontn@me}%
    }%
  \else
    \errmessage{Family `#1' already defined}%
  \fi
}

%                symbol         type fam    pos (hex)
% \NewMathSymbol {\blacksquare} {0}  {AMSa} {04}

\def\NewMathSymbol#1#2#3#4{%
  \edef\f@mly{\expandafter\hexnumber{\csname sym#3fam\endcsname}}%
  \mathchardef#1="#2\f@mly#4\relax
}

%                  macro name  type  fam1   pos  fam2   pos
% \NewMathDelimiter{\ulcorner} {4}   {AMSa} {70} {AMSb} {70}

\newif\ifd@f

\def\NewMathDelimiter#1#2#3#4#5#6{%
  \d@ftrue
  \expandafter\ifx\csname sym#3fam\endcsname\relax
    \d@ffalse \errmessage{Family `#3' is not defined}%
  \fi
  \expandafter\ifx\csname sym#5fam\endcsname\relax
    \d@ffalse \errmessage{Family `#5' is not defined}%
  \fi
  \ifd@f
    \edef\f@mly{\expandafter\hexnumber{\csname sym#3fam\endcsname}}%
    \edef\f@mlytw@{\expandafter\hexnumber{\csname sym#5fam\endcsname}}%
    \xdef#1{\delimiter"#2\f@mly #4\f@mlytw@ #6\relax}%
  \fi
}

%                  macro name  base font  skewchar setting e.g '60 (octal)
% \NewMathAlphabet {mathbssi}  {mtmisb10} {}

\def\setboxz@h{\setbox\z@\hbox}
\def\wdz@{\wd\z@}
\def\boxz@{\box\z@}
\def\setbox@ne{\setbox\@ne}
\def\wd@ne{\wd\@ne}

\def\math@atom#1#2{%
   \binrel@{#1}\binrel@@{#2}}
\def\binrel@#1{\setboxz@h{\thinmuskip0mu
  \medmuskip\m@ne mu\thickmuskip\@ne mu$#1\m@th$}%
 \setbox@ne\hbox{\thinmuskip0mu\medmuskip\m@ne mu\thickmuskip
  \@ne mu${}#1{}\m@th$}%
 \setbox\tw@\hbox{\hskip\wd@ne\hskip-\wdz@}}
\def\binrel@@#1{\ifdim\wd2<\z@\mathbin{#1}\else\ifdim\wd\tw@>\z@
 \mathrel{#1}\else{#1}\fi\fi}

\def\m@thit{1}

\def\set@skchar#1{\global\expandafter\skewchar
  \csname\fontn@me\endcsname=#1\relax}

\def\NewMathAlphabet#1#2#3{%
  \def\tst{#3}%
  \ifx\tst\empty\else % if a \skewchar setting is present
    \expandafter\gdef\csname #1@sc\endcsname{}%  \def\cmd@sc{..}
  \fi
  \expandafter\def\csname #1\endcsname{%  \def\cmd{\protect\@cmd}
    \protect\csname @#1\endcsname}%
  \expandafter\def\csname @#1\endcsname##1{%  \def\@cmd{..}
    {%
    \begingroup
      \get@font{#1}{#2}{\s@ze}%
      \@ifundefined{#1@sc}{}{\set@skchar{#3}}%
      \ass@tfont{m@thit}{\fontn@me}%
      \get@font{#1}{#2}{\ss@ze}%
      \@ifundefined{#1@sc}{}{\set@skchar{#3}}%
      \ass@sfont{m@thit}{\fontn@me}%
      \get@font{#1}{#2}{\sss@ze}%
      \@ifundefined{#1@sc}{}{\set@skchar{#3}}%
      \ass@ssfont{m@thit}{\fontn@me}%
      \math@atom{##1}{%
      \mathchoice%
        {\hbox{$\m@th\displaystyle##1$}}%
        {\hbox{$\m@th\textstyle##1$}}%
        {\hbox{$\m@th\scriptstyle##1$}}%
        {\hbox{$\m@th\scriptscriptstyle##1$}}}%
    \endgroup
    }%
  }%
}

%                  macro name  base font  hyphenchar setting e.g -1 (off)
% \NewTextAlphabet {textbfit}  {mtbxti10} {}

% save a family if \NewTextAlphabet is not used.
\newif\iffirstta  \firsttatrue

\def\set@hchar#1{\global\expandafter\hyphenchar
  \csname\fontn@me\endcsname=#1\relax}

\def\NewTextAlphabet#1#2#3{%
  \iffirstta
    \global\firsttafalse
    \newfam\scratchfam
    \edef\scrt@fam{\the\allocationnumber}%
  \fi
  \def\tst{#3}%
  \ifx\tst\empty\else % if a \hyphenchar setting is required
    \expandafter\gdef\csname #1@hc\endcsname{}%  \def\cmd@sc{..}
  \fi
  \expandafter\def\csname #1\endcsname{%  \def\cmd{\protect\t@cmd}
    \protect\csname t@#1\endcsname}%
  \long\expandafter\def\csname t@#1\endcsname##1{%  \def\t@cmd{..}
    \ifmmode
      \typeout{Warning: do not use \expandafter\string\csname #1\endcsname
        \space in math mode}\fi%
    {%
      \get@font{#1}{#2}{\s@ze}\let\t@xtfnt=\fontn@me\relax
      \@ifundefined{#1@hc}{}{\set@hchar{#3}}%
      \ass@tfont{scrt@fam}{\fontn@me}%
      \get@font{#1}{#2}{\ss@ze}%
      \@ifundefined{#1@hc}{}{\set@hchar{#3}}%
      \ass@sfont{scrt@fam}{\fontn@me}%
      \get@font{#1}{#2}{\sss@ze}%
      \@ifundefined{#1@hc}{}{\set@hchar{#3}}%
      \ass@ssfont{scrt@fam}{\fontn@me}%
      \fam\scratchfam\csname\t@xtfnt\endcsname
    ##1%
    }%
  }%
  \expandafter\def\csname #1shape%  \def\cmdshape{\protect\@cmdshape}
    \endcsname{\protect\csname @#1shape\endcsname}%
  \expandafter\def\csname @#1shape\endcsname{%  \def\@cmdshape
    \ifmmode
      \typeout{Warning: do not use \expandafter\string\csname
        #1shape\endcsname \space in math mode}\fi
      \get@font{#1}{#2}{\s@ze}\let\t@xtfnt=\fontn@me\relax
      \@ifundefined{#1@hc}{}{\set@hchar{#3}}%
      \ass@tfont{scrt@fam}{\fontn@me}%
      \get@font{#1}{#2}{\ss@ze}%
      \@ifundefined{#1@hc}{}{\set@hchar{#3}}%
      \ass@sfont{scrt@fam}{\fontn@me}%
      \get@font{#1}{#2}{\sss@ze}%
      \@ifundefined{#1@hc}{}{\set@hchar{#3}}%
      \ass@ssfont{scrt@fam}{\fontn@me}%
      \fam\scratchfam\csname\t@xtfnt\endcsname
  }%
}

% \bmath{math text}

\ifprod@font
  \def\math@itfnt{mtmib10}
  \def\math@syfnt{mtbsy10}
\else
  \def\math@itfnt{cmmib10}
  \def\math@syfnt{cmbsy10}
\fi

\def\m@thsy{2}

\def\bmath{\protect\@bmath}
\def\@bmath#1{%
  {%
  \begingroup
    \get@font{mthit}{\math@itfnt}{\s@ze}\set@skchar{'177}%
    \ass@tfont{m@thit}{\fontn@me}%
    \get@font{mthit}{\math@itfnt}{\ss@ze}\set@skchar{'177}%
    \ass@sfont{m@thit}{\fontn@me}%
    \get@font{mthit}{\math@itfnt}{\sss@ze}\set@skchar{'177}%
    \ass@ssfont{m@thit}{\fontn@me}%
    \get@font{mthsy}{\math@syfnt}{\s@ze}\set@skchar{'60}%
    \ass@tfont{m@thsy}{\fontn@me}%
    \get@font{mthsy}{\math@syfnt}{\ss@ze}\set@skchar{'60}%
    \ass@sfont{m@thsy}{\fontn@me}%
    \get@font{mthsy}{\math@syfnt}{\sss@ze}\set@skchar{'60}%
    \ass@ssfont{m@thsy}{\fontn@me}%
    \math@atom{#1}{%
    \mathchoice%
      {\hbox{$\m@th\displaystyle#1$}}%
      {\hbox{$\m@th\textstyle#1$}}%
      {\hbox{$\m@th\scriptstyle#1$}}%
      {\hbox{$\m@th\scriptscriptstyle#1$}}}%
  \endgroup
  }%
}

%%%%%%%%%%%%%%%%%%%%%%%%%%%%%%%%%%%%%%%%%

% Astronomy and Astrophysics symbol macros

\def\la{\mathrel{\mathchoice {\vcenter{\offinterlineskip\halign{\hfil
$\displaystyle##$\hfil\cr<\cr\sim\cr}}}
{\vcenter{\offinterlineskip\halign{\hfil$\textstyle##$\hfil\cr
<\cr\sim\cr}}}
{\vcenter{\offinterlineskip\halign{\hfil$\scriptstyle##$\hfil\cr
<\cr\sim\cr}}}
{\vcenter{\offinterlineskip\halign{\hfil$\scriptscriptstyle##$\hfil\cr
<\cr\sim\cr}}}}}

\def\ga{\mathrel{\mathchoice {\vcenter{\offinterlineskip\halign{\hfil
$\displaystyle##$\hfil\cr>\cr\sim\cr}}}
{\vcenter{\offinterlineskip\halign{\hfil$\textstyle##$\hfil\cr
>\cr\sim\cr}}}
{\vcenter{\offinterlineskip\halign{\hfil$\scriptstyle##$\hfil\cr
>\cr\sim\cr}}}
{\vcenter{\offinterlineskip\halign{\hfil$\scriptscriptstyle##$\hfil\cr
>\cr\sim\cr}}}}}

\def\lid{\mathrel{\mathchoice {\vcenter{\offinterlineskip\halign{\hfil
$\displaystyle##$\hfil\cr<\cr\noalign{\vskip1.2pt}=\cr}}}
{\vcenter{\offinterlineskip\halign{\hfil$\textstyle##$\hfil\cr<\cr
\noalign{\vskip1.2pt}=\cr}}}
{\vcenter{\offinterlineskip\halign{\hfil$\scriptstyle##$\hfil\cr<\cr
\noalign{\vskip1pt}=\cr}}}
{\vcenter{\offinterlineskip\halign{\hfil$\scriptscriptstyle##$\hfil\cr
<\cr
\noalign{\vskip0.9pt}=\cr}}}}}

\def\degr{\hbox{$^\circ$}}
\def\diameter{{\ifmmode\mathchoice
{\ooalign{\hfil\hbox{$\displaystyle/$}\hfil\crcr
{\hbox{$\displaystyle\mathchar"20D$}}}}
{\ooalign{\hfil\hbox{$\textstyle/$}\hfil\crcr
{\hbox{$\textstyle\mathchar"20D$}}}}
{\ooalign{\hfil\hbox{$\scriptstyle/$}\hfil\crcr
{\hbox{$\scriptstyle\mathchar"20D$}}}}
{\ooalign{\hfil\hbox{$\scriptscriptstyle/$}\hfil\crcr
{\hbox{$\scriptscriptstyle\mathchar"20D$}}}}
\else{\ooalign{\hfil/\hfil\crcr\mathhexbox20D}}%
\fi}}

\def\sq{\ifmmode\squareforqed\else{\unskip\nobreak\hfil
\penalty50\hskip1em\null\nobreak\hfil\squareforqed
\parfillskip=0pt\finalhyphendemerits=0\endgraf}\fi}
\def\squareforqed{\hbox{\rlap{$\sqcap$}$\sqcup$}}

\def\fs{\hbox{$.\!\!^{\rm s}$}}

\def\farcm{\hbox{$.\mkern-4mu^\prime$}}
\def\farcs{\hbox{$.\!\!^{\prime\prime}$}}

\def\arcmin{\hbox{$^\prime$}}

% Simulated Blackboard Bold symbols

\def\bbbf{{\rm I\!F}}
\def\bbbh{{\rm I\!H}}

\def\bbbp{{\rm I\!P}}

\def\bbbc{{\mathchoice {\setbox0=\hbox{$\displaystyle\rm C$}\hbox{\hbox
to0pt{\kern0.4\wd0\vrule height0.9\ht0\hss}\box0}}
{\setbox0=\hbox{$\textstyle\rm C$}\hbox{\hbox
to0pt{\kern0.4\wd0\vrule height0.9\ht0\hss}\box0}}
{\setbox0=\hbox{$\scriptstyle\rm C$}\hbox{\hbox
to0pt{\kern0.4\wd0\vrule height0.9\ht0\hss}\box0}}
{\setbox0=\hbox{$\scriptscriptstyle\rm C$}\hbox{\hbox
to0pt{\kern0.4\wd0\vrule height0.9\ht0\hss}\box0}}}}
\def\bbbq{{\mathchoice {\setbox0=\hbox{$\displaystyle\rm
Q$}\hbox{\raise
0.15\ht0\hbox to0pt{\kern0.4\wd0\vrule height0.8\ht0\hss}\box0}}
{\setbox0=\hbox{$\textstyle\rm Q$}\hbox{\raise
0.15\ht0\hbox to0pt{\kern0.4\wd0\vrule height0.8\ht0\hss}\box0}}
{\setbox0=\hbox{$\scriptstyle\rm Q$}\hbox{\raise
0.15\ht0\hbox to0pt{\kern0.4\wd0\vrule height0.7\ht0\hss}\box0}}
{\setbox0=\hbox{$\scriptscriptstyle\rm Q$}\hbox{\raise
0.15\ht0\hbox to0pt{\kern0.4\wd0\vrule height0.7\ht0\hss}\box0}}}}
\def\bbbt{{\mathchoice {\setbox0=\hbox{$\displaystyle\rm
T$}\hbox{\hbox to0pt{\kern0.3\wd0\vrule height0.9\ht0\hss}\box0}}
{\setbox0=\hbox{$\textstyle\rm T$}\hbox{\hbox
to0pt{\kern0.3\wd0\vrule height0.9\ht0\hss}\box0}}
{\setbox0=\hbox{$\scriptstyle\rm T$}\hbox{\hbox
to0pt{\kern0.3\wd0\vrule height0.9\ht0\hss}\box0}}
{\setbox0=\hbox{$\scriptscriptstyle\rm T$}\hbox{\hbox
to0pt{\kern0.3\wd0\vrule height0.9\ht0\hss}\box0}}}}
\def\bbbs{{\mathchoice
{\setbox0=\hbox{$\displaystyle     \rm S$}\hbox{\raise0.5\ht0\hbox
to0pt{\kern0.35\wd0\vrule height0.45\ht0\hss}\hbox
to0pt{\kern0.55\wd0\vrule height0.5\ht0\hss}\box0}}
{\setbox0=\hbox{$\textstyle        \rm S$}\hbox{\raise0.5\ht0\hbox
to0pt{\kern0.35\wd0\vrule height0.45\ht0\hss}\hbox
to0pt{\kern0.55\wd0\vrule height0.5\ht0\hss}\box0}}
{\setbox0=\hbox{$\scriptstyle      \rm S$}\hbox{\raise0.5\ht0\hbox
to0pt{\kern0.35\wd0\vrule height0.45\ht0\hss}\raise0.05\ht0\hbox
to0pt{\kern0.5\wd0\vrule height0.45\ht0\hss}\box0}}
{\setbox0=\hbox{$\scriptscriptstyle\rm S$}\hbox{\raise0.5\ht0\hbox
to0pt{\kern0.4\wd0\vrule height0.45\ht0\hss}\raise0.05\ht0\hbox
to0pt{\kern0.55\wd0\vrule height0.45\ht0\hss}\box0}}}}
\def\bbbz{{\mathchoice {\hbox{$\sf\textstyle Z\kern-0.4em Z$}}
{\hbox{$\sf\textstyle Z\kern-0.4em Z$}}
{\hbox{$\sf\scriptstyle Z\kern-0.3em Z$}}
{\hbox{$\sf\scriptscriptstyle Z\kern-0.2em Z$}}}}

% NUMBER THE DESIGN ELEMENTS

\def\Nulle{0} % null element
\def\Afe{1}   % author affiliation
\def\Hae{2}   % heading A
\def\Hbe{3}   % heading B
\def\Hce{4}   % heading C
\def\Hde{5}   % heading D

% TEMPORARY REGISTERS

\newcount\LastMac       \LastMac=\Nulle

\newskip\half      \half=5.5pt plus 1.5pt minus 2.25pt
\newskip\one       \one=11pt plus 3pt minus 5.5pt
\newskip\onehalf   \onehalf=16.5pt plus 5.5pt minus 8.25pt
\newskip\two       \two=22pt plus 5.5pt minus 11pt

\def\Half{\addvspace{\half}}
\def\One{\addvspace{\one}}
\def\OneHalf{\addvspace{\onehalf}}
\def\Two{\addvspace{\two}}

\def\Raggedright{% set lines unjustified
  \rightskip=\z@ plus \hsize\relax
}

\def\Fullout{% set lines justified
  \rightskip=\z@\relax
}

\def\Hang#1#2{% set hanging indentation
  \hangindent=#1%
  \hangafter=#2\relax
}

% Pagestyles

\newif\ifsp@page
\def\pagestyle#1{\csname ps@#1\endcsname}
\def\thispagestyle#1{\global\sp@pagetrue\gdef\sp@type{#1}}

\def\ps@titlepage{%
  \def\@oddhead{\eightpoint\noindent \the\CatchLine
    \ifprod@font\else\qquad Printed\ \today\qquad
      (MN plain \TeX\ macros\ v\@version)\fi \hfil}%
  \let\@evenhead=\@oddhead
  \def\@oddfoot{\eightpoint\copyright\ \@pubyear\ RAS\hfil}%
  \def\@evenfoot{\hfil\eightpoint\noindent\copyright\ \@pubyear\ RAS}%
}

\def\ps@headings{%
  \def\@oddhead{\elevenpoint\it\noindent
    \hfill\the\RightHeader\hskip1.5em\rm\folio}%
  \def\@evenhead{\elevenpoint\noindent
    \folio\hskip1.5em\it\the\LeftHeader\hfill}%
  \def\@oddfoot{\eightpoint\noindent\copyright\ \@pubyear\ RAS,
    MNRAS {\bf \@volume}, \@pagerange\hfil}%
  \def\@evenfoot{\hfil\eightpoint\copyright\ \@pubyear\ RAS,
    MNRAS {\bf \@volume}, \@pagerange}%
}

\def\ps@plate{%
  \def\@oddhead{\eightpoint\noindent\plt@cap\hfil}%
  \def\@evenhead{\eightpoint\noindent\plt@cap\hfil}%
  \def\@oddfoot{\eightpoint\noindent\copyright\ \@pubyear\ RAS,
    MNRAS {\bf \@volume}, \@pagerange\hfil}%
  \def\@evenfoot{\hfil\eightpoint\copyright\ \@pubyear\ RAS,
    MNRAS {\bf \@volume}, \@pagerange}%
}

% DESIGN ELEMENT DEFINITIONS

% Article opening

\def\title#1{% article title
  \bgroup
    \vbox to 8pt{\vss}%
    \seventeenpoint
    \Raggedright
    \noindent \strut{\bf #1}\par
  \egroup
}

\def\author#1{% article author(s)
  \bgroup
    \ifnum\LastMac=\Afe \OneHalf\else \vskip 21pt\fi
    \fourteenpoint
    \Raggedright
    \noindent \strut #1\par
    \vskip 3pt%
  \egroup
}

\def\affiliation#1{% author(s) affiliation
  \bgroup
    \vskip -4pt%
    \eightpoint
    \Raggedright
    \noindent \strut {\it #1}\par
  \egroup
  \LastMac=\Afe\relax
}

\def\acceptedline#1{% acceptance date
  \bgroup
    \Two
    \eightpoint
    \Raggedright
    \noindent \strut #1\par
  \egroup
}

\long\def\abstract#1{%
  \bgroup
    \vskip 20pt%
    \leftskip 11pc\rightskip\z@
    \noindent{\ninebf ABSTRACT}\par
    \tenpoint
    \Fullout
    \noindent #1\par
  \egroup
}

\long\def\keywords#1{% keywords
  \bgroup
    \Half
    \leftskip 11pc\rightskip\z@
    \tenpoint
    \Fullout
    \noindent\hbox{\bf Key words:}\ #1\par
  \egroup
}

% The \maketitle macro ensures that the two spanning material appears
% at the top of the first page, and that it has two lines of space
% underneath it. If you forget this in you input, no output will be produced.
% The \BeginOpening (alias \begintopmatter) macro should be called at the
% very start of the input file, so that it is in force when the document
% starts. This ensures that when \maketitle is called that the group is
% closed, and the material gets printed.

\def\maketitle{%
  \EndOpening
  \ifsinglecol \else \MakePage\fi
}

% Page offset

% Counter setup

\def\@nameuse#1{\csname #1\endcsname}
\def\arabic#1{\@arabic{\@nameuse{#1}}}
\def\alph#1{\@alph{\@nameuse{#1}}}
\def\Alph#1{\@Alph{\@nameuse{#1}}}
\def\@arabic#1{\number #1}
\def\@Alph#1{\ifcase#1\or A\or B\or C\or D\else\@Ialph{#1}\fi}
\def\@Ialph#1{\ifcase#1\or \or \or \or \or E\or F\or G\or H\or I\or J\or
   K\or L\or M\or N\or O\or P\or Q\or R\or S\or T\or U\or V\or W\or X\or
   Y\or Z\else\errmessage{Counter out of range}\fi}
\def\@alph#1{\ifcase#1\or a\or b\or c\or d\else\@ialph{#1}\fi}
\def\@ialph#1{\ifcase#1\or \or \or \or \or e\or f\or g\or h\or i\or j\or
   k\or l\or m\or n\or o\or p\or q\or r\or s\or t\or u\or v\or w\or x\or y\or
   z\else\errmessage{Counter out of range}\fi}

% Equation auto-numbering

\newcount\Eqnno
\newcount\SubEqnno

\def\theeq{\arabic{Eqnno}}
\def\thesubeq{\alph{SubEqnno}}

\def\stepeq{\relax
  \global\SubEqnno \z@
  \global\advance\Eqnno \@ne\relax
  {\rm (\theeq)}%
}

\def\startsubeq{\relax
  \global\SubEqnno \z@
  \global\advance\Eqnno \@ne\relax
  \stepsubeq
}

\def\stepsubeq{\relax
  \global\advance\SubEqnno \@ne\relax
  {\rm (\theeq\thesubeq)}%
}

% Headings

\newcount\Sec        %  heading auto number counters
\newcount\SecSec
\newcount\SecSecSec

\def\thesection{\arabic{Sec}}
\def\thesubsection{\thesection.\arabic{SecSec}}
\def\thesubsubsection{\thesubsection.\arabic{SecSecSec}}

\Sec=\z@

\def\:{\let\@sptoken= } \:  % this makes \@sptoken a space token 
\def\:{\@xifnch} \expandafter\def\: {\futurelet\@tempc\@ifnch}

\def\@ifnextchar#1#2#3{%
  \let\@tempMACe #1%
  \def\@tempMACa{#2}%
  \def\@tempMACb{#3}%
  \futurelet \@tempMACc\@ifnch%
}

\def\@ifnch{%
\ifx \@tempMACc \@sptoken%
  \let\@tempMACd\@xifnch%
\else%
  \ifx \@tempMACc \@tempMACe%
    \let\@tempMACd\@tempMACa%
  \else%
    \let\@tempMACd\@tempMACb%
  \fi%
\fi%
\@tempMACd%
}

\def\@ifstar#1#2{\@ifnextchar *{\def\@tempMACa*{#1}\@tempMACa}{#2}}

\newskip\@tempskipb

\def\addvspace#1{%
  \ifvmode\else \endgraf\fi%
  \ifdim\lastskip=\z@%
    \vskip #1\relax%
  \else%
    \@tempskipb#1\relax\@xaddvskip%
  \fi%
}

\def\@xaddvskip{%
  \ifdim\lastskip<\@tempskipb%
    \vskip-\lastskip%
    \vskip\@tempskipb\relax%
  \else%
    \ifdim\@tempskipb<\z@%
      \ifdim\lastskip<\z@ \else%
        \advance\@tempskipb\lastskip%
        \vskip-\lastskip\vskip\@tempskipb%
      \fi%
    \fi%
  \fi%
}

\newskip\@tmpSKIP

\def\addpen#1{%
  \ifvmode
    \if@nobreak
    \else
      \ifdim\lastskip=\z@
        \penalty#1\relax
      \else
        \@tmpSKIP=\lastskip
        \vskip -\lastskip
        \penalty#1\vskip\@tmpSKIP
      \fi
    \fi
  \fi
}

\newcount\@clubpen   \@clubpen=\clubpenalty
\newif\if@nobreak    \@nobreakfalse

\def\@noafterindent{%
  \global\@nobreaktrue
  \everypar{\if@nobreak
              \global\@nobreakfalse
              \clubpenalty \@M
              {\setbox\z@\lastbox}%
              \LastMac=\Nulle\relax%
            \else
              \clubpenalty \@clubpen
              \everypar{}%
            \fi}%
}

\newcount\gds@cbrk   \gds@cbrk=-300

\def\@nohdbrk{\interlinepenalty \@M\relax}

\let\@par=\par
\def\@restorepar{\def\par{\@par}}

\newif\if@endpe   \@endpefalse
 
\def\@doendpe{\@endpetrue \@nobreakfalse \LastMac=\Nulle\relax%
     \def\par{\@restorepar\everypar{}\par\@endpefalse}%
              \everypar{\setbox\z@\lastbox\everypar{}\@endpefalse}%
}

\def\section{\@ifstar{\@ssection}{\@section}}

\def\@section#1{% heading A (\section{....})
  \if@nobreak
    \everypar{}%
    \ifnum\LastMac=\Hae \addvspace{\half}\fi
  \else
    \addpen{\gds@cbrk}%
    \addvspace{\two}%
  \fi
  \bgroup
    \ninepoint\bf
    \Raggedright
    \global\advance\Sec \@ne
    \ifappendix
      \global\Eqnno=\z@ \global\SubEqnno=\z@\relax
      \def\ch@ck{#1}%
      \ifx\ch@ck\empty \def\c@lon{}\else\def\c@lon{:}\fi
      \noindent\@nohdbrk APPENDIX\ \thesection\c@lon\hskip 0.5em%
        \uppercase{#1}\par
    \else
      \noindent\@nohdbrk\thesection\hskip 1pc \uppercase{#1}\par
    \fi
    \global\SecSec=\z@
  \egroup
  \nobreak
  \vskip\half
  \nobreak
  \@noafterindent
  \LastMac=\Hae\relax
}

\def\@ssection#1{%  main section heading (\section*{....})
  \if@nobreak
    \everypar{}%
    \ifnum\LastMac=\Hae \addvspace{\half}\fi
  \else
    \addpen{\gds@cbrk}%
    \addvspace{\two}%
  \fi
  \bgroup
    \ninepoint\bf
    \Raggedright
%    \ifappendix
%      \global\Eqnno=\z@ \global\SubEqnno=\z@\relax % mh in apps dont reset
%      \noindent\@nohdbrk APPENDIX:\hskip 0.5em%
%        \uppercase{#1}\par
%    \else
    \noindent\@nohdbrk\uppercase{#1}\par
%    \fi
  \egroup
  \nobreak
  \vskip\half
  \nobreak
  \@noafterindent
  \LastMac=\Hae\relax
}

\def\subsection{\@ifstar{\@ssubsection}{\@subsection}}

\def\@subsection#1{% heading B
  \if@nobreak
    \everypar{}%
    \ifnum\LastMac=\Hae \addvspace{1pt plus 1pt minus .5pt}\fi
  \else
    \addpen{\gds@cbrk}%
    \addvspace{\onehalf}%
  \fi
  \bgroup
    \ninepoint\bf
    \Raggedright
    \global\advance\SecSec \@ne
    \noindent\@nohdbrk\thesubsection \hskip 1pc\relax #1\par
    \global\SecSecSec=\z@
  \egroup
  \nobreak
  \vskip\half
  \nobreak
  \@noafterindent
  \LastMac=\Hbe\relax
}

\def\@ssubsection#1{% heading B*
  \if@nobreak
    \everypar{}%
    \ifnum\LastMac=\Hae \addvspace{1pt plus 1pt minus .5pt}\fi
  \else
    \addpen{\gds@cbrk}%
    \addvspace{\onehalf}%
  \fi
  \bgroup
    \ninepoint\bf
    \Raggedright
    \noindent\@nohdbrk #1\par
  \egroup
  \nobreak
  \vskip\half
  \nobreak
  \@noafterindent
  \LastMac=\Hbe\relax
}

\def\subsubsection{\@ifstar{\@ssubsubsection}{\@subsubsection}}

\def\@subsubsection#1{% heading C
  \if@nobreak
    \everypar{}%
    \ifnum\LastMac=\Hbe \addvspace{1pt plus 1pt minus .5pt}\fi
  \else
    \addpen{\gds@cbrk}%
    \addvspace{\onehalf}%
  \fi
  \bgroup
    \ninepoint\it
    \Raggedright
    \global\advance\SecSecSec \@ne
    \noindent\@nohdbrk\thesubsubsection \hskip 1pc\relax #1\par
  \egroup
  \nobreak
  \vskip\half
  \nobreak
  \@noafterindent
  \LastMac=\Hce\relax
}

\def\@ssubsubsection#1{% heading C*
  \if@nobreak
    \everypar{}%
    \ifnum\LastMac=\Hbe \addvspace{1pt plus 1pt minus .5pt}\fi
  \else
    \addpen{\gds@cbrk}%
    \addvspace{\onehalf}%
  \fi
  \bgroup
    \ninepoint\it
    \Raggedright
    \noindent\@nohdbrk #1\par
  \egroup
  \nobreak
  \vskip\half
  \nobreak
  \@noafterindent
  \LastMac=\Hce\relax
}

\def\paragraph#1{% heading D
  \if@nobreak
    \everypar{}%
  \else
    \addpen{\gds@cbrk}%
    \addvspace{\one}%
  \fi%
  \bgroup%
    \ninepoint\it
    \noindent #1\ \nobreak%
  \egroup
  \LastMac=\Hde\relax
  \ignorespaces
}

% Appendix

\newif\ifappendix

\def\appendix{%
  \global\appendixtrue
  \def\thesection{\Alph{Sec}}%
  \def\thesubsection{\thesection\arabic{SecSec}}%
  \def\theeq{\thesection\arabic{Eqnno}}%
  \Sec=\z@ \SecSec=\z@ \SecSecSec=\z@ \Eqnno=\z@ \SubEqnno=\z@\relax
}

% Text

 % provided for backward compatibility

% Lists

\def\beginlist{%
  \par\if@nobreak \else\addvspace{\half}\fi%
  \bgroup%
    \ninepoint
    \let\item=\list@item%
}

\def\list@item{%
  \par\noindent\hskip 1em\relax%
  \ignorespaces%
}

\def\endlist{\par\egroup\addvspace{\half}\@doendpe}

% References

\def\beginrefs{%
  \par
  \bgroup
    \eightpoint
    \Fullout
    \let\bibitem=\bib@item
}

\def\bib@item{%
  \par\parindent=1.5em\Hang{1.5em}{1}%
  \everypar={\Hang{1.5em}{1}\ignorespaces}%
  \noindent\ignorespaces
}

\def\endrefs{\par\egroup\@doendpe}

% Page heads

\newtoks\CatchLine

\def\@journal{Mon.\ Not.\ R.\ Astron.\ Soc.\ }  % The journal title string
\def\@pubyear{1994}        % Assign a default publication year
\def\@pagerange{000--000}  % Assign a default page-range
\def\@volume{000}          % Assign a default volume number
\def\@microfiche{}         %

\def\pubyear#1{\gdef\@pubyear{#1}\@makecatchline}
\def\pagerange#1{\gdef\@pagerange{#1}\@makecatchline}
\def\volume#1{\gdef\@volume{#1}\@makecatchline}
\def\microfiche#1{\gdef\@microfiche{and Microfiche\ #1}\@makecatchline}

\def\@makecatchline{%
  \global\CatchLine{%
    {\rm \@journal {\bf \@volume},\ \@pagerange\ (\@pubyear)\ \@microfiche}}%
}

\@makecatchline % Assign a catchline, using the above defaults

\newtoks\LeftHeader
\def\shortauthor#1{% left page head
  \global\LeftHeader{#1}%
}

\newtoks\RightHeader
\def\shorttitle#1{% right page head
  \global\RightHeader{#1}%
}

\def\PageHead{% recto/verso running heads
  \begingroup
    \ifsp@page
      \csname ps@\sp@type\endcsname
    \fi
    \ifodd\pageno
      \let\the@head=\@oddhead
    \else
      \let\the@head=\@evenhead
    \fi
    \vbox to \z@{\vskip-22.5\p@%
      \hbox to \PageWidth{\vbox to8.5\p@{}%
        \the@head
      }%
    \vss}%
  \endgroup
  \nointerlineskip
}

\gdef\PageFoot{%
  \nointerlineskip%
  \begingroup
  \ifsp@page
    \csname ps@\sp@type\endcsname
    \global\sp@pagefalse
  \fi
  \vbox to 22pt{\vfil%
    \hbox to \PageWidth{%
      \eightpoint\strut\noindent
      \ifodd\pageno
        \@oddfoot
      \else
        \@evenfoot
      \fi
    }%
  }%
  \endgroup
}

\def\today{%
  \number\day\space
  \ifcase\month\or January\or February\or March\or April\or May\or June\or
    July\or August\or September\or October\or November\or December\fi
  \space\number\year%
}

\def\authorcomment#1{%
  \gdef\PageFoot{%
    \nointerlineskip%
    \vbox to 20pt{\vfil%
      \hbox to \PageWidth{\elevenpoint\noindent \hfil #1 \hfil}}%
  }%
}

% Plate pages

\newif\ifplate@page
\newbox\plt@box

\def\beginplatepage{%
  \let\plate=\plate@head
  \let\caption=\fig@caption
  \global\setbox\plt@box=\vbox\bgroup
  \TEMPDIMEN=\PageWidth % For \fig@caption test
  \hsize=\PageWidth\relax
}

\def\endplatepage{\par\egroup\global\plate@pagetrue}
\def\plate@head#1{\gdef\plt@cap{#1}}

% Letters option

\def\letters{%
  \gdef\folio{\ifnum\pageno<\z@ L\romannumeral-\pageno
    \else L\number\pageno \fi}%
}

% Math setup

% The standard math indentation
\newdimen\mathindent

\global\mathindent=\z@
\global\everydisplay{\global\@dspwd=\displaywidth\displaysetup}

% New versions of \displaylines, \eqalign, \eqalignno for
% when non-centered math is in use.

\def\@displaylines#1{% (for non-centered math)
  {}$\displ@y\hbox{\vbox{\halign{$\@lign\hfil\displaystyle##\hfil$\crcr
  #1\crcr}}}${}%
}

\def\@eqalign#1{\null\vcenter{\openup\jot\m@th% (for non-centered math)
  \ialign{\strut\hfil$\displaystyle{##}$&$\displaystyle{{}##}$\hfil
      \crcr#1\crcr}}%
}

\def\@eqalignno#1{% (for non-centered math)
  \global\advance\@dspwd by -\mathindent%
  {}$\displ@y\hbox{\vbox{\halign to\@dspwd%
  {\hfil$\@lign\displaystyle{##}$\tabskip\z@skip
  &$\@lign\displaystyle{{}##}$\hfil\tabskip\centering
  &\llap{$\@lign##$}\tabskip\z@skip\crcr
  #1\crcr}}}${}%
}

% When equations are flushleft ensure, that \displaylines,
% \eqalign, \eqalignno and \leqalignno point to the new versions of
% the macros. Also make \leqalignno act like \eqalignno, otherwise the
% equation text would `crash' into the equation number.

\global\let\displaylines=\@displaylines
\global\let\eqalign=\@eqalign
\global\let\eqalignno=\@eqalignno
\global\let\leqalignno=\@eqalignno

\newdimen\@dspwd   \@dspwd=\z@
\newif\if@eqno
\newif\if@leqno
\newtoks\@eqn
\newtoks\@eq

\def\displaysetup#1$${\displaytest#1\eqno\eqno\displaytest}

\def\displaytest#1\eqno#2\eqno#3\displaytest{%
 \if!#3!\ldisplaytest#1\leqno\leqno\ldisplaytest
 \else\@eqnotrue\@leqnofalse\@eqn={#2}\@eq={#1}\fi
 \generaldisplay$$}

\def\ldisplaytest#1\leqno#2\leqno#3\ldisplaytest{%
\@eq={#1}%
 \if!#3!\@eqnofalse\else\@eqnotrue\@leqnotrue
  \@eqn={#2}\fi}

\def\generaldisplay{%
  \if@eqno
    \if@leqno
      \hbox to \displaywidth{\noindent
        \rlap{$\displaystyle\the\@eqn$}%
        \hskip\mathindent$\displaystyle\the\@eq$\hfil}%
    \else
      \hbox to \displaywidth{\noindent
        \hskip\mathindent
        $\displaystyle\the\@eq$\hfil$\displaystyle\the\@eqn$}%
    \fi
  \else
    \hbox to \displaywidth{\noindent
      \hskip\mathindent$\displaystyle\the\@eq$\hfil}%
  \fi
}

% Finishing notice

\def\@notice{%
  \par\Two%
  \noindent{\b@ls{11pt}\ninerm This paper has been produced using the
    Royal Astronomical Society/Blackwell Science \TeX\ macros.\par}%
}

% redefine \bye to output our identification notice :
\outer\def
\section*{References}
\beginrefs
\bibitem Aghanim N., Desert F. X., Puget J. L., Gispert R., 1996, A\&A, 311, 1
\bibitem Bennett C. L. et al., 1996, ApJ, 464, L1
\bibitem Bond J. R., Myers S. T., 1996, ApJS, 103, 63
\bibitem Church S. E., Ganga K. M., Ade P. A. R., Holzapfel W. L., 
         Mauskopf P. D., Wilbanks T. M., Lange A. E., 1997, 
         astro-ph 9702196
\bibitem Ganga K., Ratra B., Church S. E., Sugiyama N., Ade P. A. R.,
         Holzapfel W. L., Lange A. E., Mauskopf P. D., 1997,
	 astro-ph 9702186
\bibitem Gough R., 1994, Australia Telescope 
         Technical Document \hfil\break AT/39.2/063
\bibitem Hattori M. et al., 1997, Nature, 388, 146
\bibitem Hu W., Scott D., Silk J., 1994, Phys. Rev. D, 49, 2
\bibitem Hu W., Sugiyama N., 1994, ApJ, 436, 456
\bibitem Hu W., Bunn E. F., Sugiyama N., 1995, ApJ, 447, L59
\bibitem James G. L., 1985, IREECON'85 International Digest, p.713
\bibitem James G. L., 1987, Proc. IEEE, 134, 217
\bibitem Jones M. E. et al., 1997, ApJ, 479, L1
\bibitem Lineweaver C. H., Barbosa D., 1997, astro-ph 9706077
\bibitem Partridge R. B., Richards E. A., Fomalont E. B., Kellermann K. I.,
         Windhorst R. A., 1997, ApJ, 483, 38
\bibitem Persi F. M., Spergel D. N., Cen R., Ostriker J. P., 1995, ApJ, 442, 1
\bibitem Readhead A. C. S., Lawrence C. R., Myers S. T., Sargent W. L. W.,
         Hardebeck H. E., Moffet A. T., 1989, ApJ, 346, 566
\bibitem Reynolds J., 1994, Australia Telescope 
         Technical Document \hfil\break AT/39.3/040
\bibitem Richards E. A., Fomalont E. B., Kellermann K. I., Partridge R. B.,
         Windhorst R. A., 1997, NRAO preprint NRAO-96/238 
\bibitem Subrahmanyan R., Ekers R. D., Sinclair M., Silk J., 1993, MNRAS,
         263, 416
\bibitem The Australia Telescope, 1992, Special issue of: J. Electr.
         Electron. Eng. Aust., 12, June
\endrefs
\bye{\@notice\par\vfill\supereject\end}

% define a sign on :

\def\start@mess{%
  Monthly notices of the RAS journal style (\@typeface)\space
    v\@version,\space \@verdate.%
}

\everyjob{\Warn{\start@mess}}

% Two-column macros

%--------------------------------------------------------%
%                     INITIALISATION                     %
%--------------------------------------------------------%

\newif\if@debug \@debugfalse  %  when false, only warnings displayed

\def\Print#1{\if@debug\immediate\write16{#1}\else \fi}
\def\Warn#1{\immediate\write16{#1}}
\def\wlog#1{}

\newcount\Iteration % temporary loop counter

\def\Single{0} \def\Double{1}                 % ItemSPAN's
\def\Figure{0} \def\Table{1}                  % ItemTYPE's

\def\InStack{0}  % ItemSTATUS
\def\InZoneA{1}
\def\InZoneB{2}
\def\InZoneC{3}

\newcount\TEMPCOUNT % temporary count register
\newdimen\TEMPDIMEN % temporary dimen register
\newbox\TEMPBOX     % temporary box register
\newbox\VOIDBOX     % a box which is permenately void

\newcount\LengthOfStack % number of items currently in stack
\newcount\MaxItems      % maximum number of items allowed in stack
\newcount\StackPointer
\newcount\Point         % used in calculation for generating the
                        % physical address of an item in the stack.
\newcount\NextFigure    % number of next figure to be output
\newcount\NextTable     % number of next table to be output
\newcount\NextItem      % Next item to consider by order in stack

\newcount\StatusStack   % set to point to top of STATUS stack
\newcount\NumStack      % set to point to top of NUMBER stack
\newcount\TypeStack     % set to point to top of TYPE stack
\newcount\SpanStack     % set to point to top of SPAN stack
\newcount\BoxStack      % set to point to top of BOX stack

\newcount\ItemSTATUS    % status of present item
\newcount\ItemNUMBER    % number of present item
\newcount\ItemTYPE      % type of present item
\newcount\ItemSPAN      % span of present item
\newbox\ItemBOX         % box of present item
\newdimen\ItemSIZE      % size of present item
                        % (calculated by GetItemBOX)

\newdimen\PageHeight    % vertical measure of full page
\newdimen\TextLeading   % distance between baselines of body text
\newdimen\Feathering    % amount of interline stretch
\newcount\LinesPerPage  % height of page in text lines
\newdimen\ColumnWidth   % width of 1 column of text
\newdimen\ColumnGap     % size of gap between columns
\newdimen\PageWidth     % = \ColumnWidth * 2 + \ColumnGap
\newdimen\BodgeHeight   % Bodge to align figures and tables with text
\newcount\Leading       % Set to same as \TextLeading above

\newdimen\ZoneBSize  % size of items in ZoneB
\newdimen\TextSize   % size of text in ZoneB
\newbox\ZoneABOX     % contains Zone A material
\newbox\ZoneBBOX     % contains Zone B material
\newbox\ZoneCBOX     % contains Zone C material

\newif\ifFirstSingleItem
\newif\ifFirstZoneA
\newif\ifMakePageInComplete
\newif\ifMoreFigures \MoreFiguresfalse % set true in join stack
\newif\ifMoreTables  \MoreTablesfalse  % set true in join stack

\newif\ifFigInZoneB % used to determine in which zone an item
\newif\ifFigInZoneC % will be placed based on what is in other
\newif\ifTabInZoneB % zones already for a given page.
\newif\ifTabInZoneC

\newif\ifZoneAFullPage

\newbox\MidBOX    % = LeftBOX+gap+RightBOX
\newbox\LeftBOX
\newbox\RightBOX
\newbox\PageBOX   % complete made-up page

\newif\ifLeftCOL  % flags first pass through output routine
\LeftCOLtrue

\newdimen\ZoneBAdjust

\newcount\ItemFits
\def\Yes{1}
\def\No{2}

% Setup file.

\MaxItems=15
\NextFigure=\z@        % used for article opening
\NextTable=\@ne

\BodgeHeight=6pt
\TextLeading=11pt    % baselineskip of body text
\Leading=11
\Feathering=\z@      % amount of interline stretch
\LinesPerPage=61     % number of text lines per full page -1
\topskip=\TextLeading
\ColumnWidth=20pc    % width of text columns
\ColumnGap=2pc       % gap between columns

\newskip\ItemSepamount  % space between floats
\ItemSepamount=\TextLeading plus \TextLeading minus 4pt

\parskip=\z@ plus .1pt
\parindent=18pt
\widowpenalty=\z@
\clubpenalty=10000
\tolerance=1500
\hbadness=1500
\abovedisplayskip=6pt plus 2pt minus 1pt
\belowdisplayskip=6pt plus 2pt minus 1pt
\abovedisplayshortskip=6pt plus 2pt minus 1pt
\belowdisplayshortskip=6pt plus 2pt minus 1pt

\frenchspacing

\ninepoint % start main text size

\PageHeight=682pt
\PageWidth=2\ColumnWidth
\advance\PageWidth by \ColumnGap

\pagestyle{headings}

%--------------------------------------------------------%
%                         STACKS                         %
%--------------------------------------------------------%

% THE ITEM STACK
% The item stack contains contains figures and tables
% in the order in which they appear in the article source
% code.

% allocate stack space

\newcount\DUMMY \StatusStack=\allocationnumber
\newcount\DUMMY \newcount\DUMMY \newcount\DUMMY 
\newcount\DUMMY \newcount\DUMMY \newcount\DUMMY 
\newcount\DUMMY \newcount\DUMMY \newcount\DUMMY
\newcount\DUMMY \newcount\DUMMY \newcount\DUMMY 
\newcount\DUMMY \newcount\DUMMY \newcount\DUMMY

\newcount\DUMMY \NumStack=\allocationnumber
\newcount\DUMMY \newcount\DUMMY \newcount\DUMMY 
\newcount\DUMMY \newcount\DUMMY \newcount\DUMMY 
\newcount\DUMMY \newcount\DUMMY \newcount\DUMMY 
\newcount\DUMMY \newcount\DUMMY \newcount\DUMMY 
\newcount\DUMMY \newcount\DUMMY \newcount\DUMMY

\newcount\DUMMY \TypeStack=\allocationnumber
\newcount\DUMMY \newcount\DUMMY \newcount\DUMMY 
\newcount\DUMMY \newcount\DUMMY \newcount\DUMMY 
\newcount\DUMMY \newcount\DUMMY \newcount\DUMMY 
\newcount\DUMMY \newcount\DUMMY \newcount\DUMMY 
\newcount\DUMMY \newcount\DUMMY \newcount\DUMMY

\newcount\DUMMY \SpanStack=\allocationnumber
\newcount\DUMMY \newcount\DUMMY \newcount\DUMMY 
\newcount\DUMMY \newcount\DUMMY \newcount\DUMMY 
\newcount\DUMMY \newcount\DUMMY \newcount\DUMMY 
\newcount\DUMMY \newcount\DUMMY \newcount\DUMMY 
\newcount\DUMMY \newcount\DUMMY \newcount\DUMMY

\newbox\DUMMY   \BoxStack=\allocationnumber
\newbox\DUMMY   \newbox\DUMMY \newbox\DUMMY 
\newbox\DUMMY   \newbox\DUMMY \newbox\DUMMY 
\newbox\DUMMY   \newbox\DUMMY \newbox\DUMMY 
\newbox\DUMMY   \newbox\DUMMY \newbox\DUMMY 
\newbox\DUMMY   \newbox\DUMMY \newbox\DUMMY

\def\wlog{\immediate\write\m@ne}

% \GetItemSTATUS, \GetItemNUMBER, \GetItemTYPE, \GetItemSPAN,
% \GetItemBox 
% are used to get details of a particular item from the item
% stack. The argument to each of these is the items position
% in the stack (usually \StackPointer)...not the items number.

\def\GetItemAll#1{%
 \GetItemSTATUS{#1}
 \GetItemNUMBER{#1}
 \GetItemTYPE{#1}
 \GetItemSPAN{#1}
 \GetItemBOX{#1}
}

% Note: \LeaveStack uses this routine. Do not destroy \Point
\def\GetItemSTATUS#1{%
 \Point=\StatusStack
 \advance\Point by #1
 \global\ItemSTATUS=\count\Point
}

% Note: \LeaveStack uses this routine. Do not destroy \Point
\def\GetItemNUMBER#1{%
 \Point=\NumStack
 \advance\Point by #1
 \global\ItemNUMBER=\count\Point
}

% Note: \LeaveStack uses this routine. Do not destroy \Point
\def\GetItemTYPE#1{%
 \Point=\TypeStack
 \advance\Point by #1
 \global\ItemTYPE=\count\Point
}

% Note: \LeaveStack uses this routine. Do not destroy \Point
\def\GetItemSPAN#1{%
 \Point\SpanStack
 \advance\Point by #1
 \global\ItemSPAN=\count\Point
}

% Note: \LeaveStack uses this routine. Do not destroy \Point
\def\GetItemBOX#1{%
 \Point=\BoxStack
 \advance\Point by #1
 \global\setbox\ItemBOX=\vbox{\copy\Point}
 \global\ItemSIZE=\ht\ItemBOX
 \global\advance\ItemSIZE by \dp\ItemBOX
 \TEMPCOUNT=\ItemSIZE
 \divide\TEMPCOUNT by \Leading
 \divide\TEMPCOUNT by 65536
 \advance\TEMPCOUNT \@ne
 \ItemSIZE=\TEMPCOUNT pt
 \global\multiply\ItemSIZE by \Leading
}

% item joins stack

\def\JoinStack{%
 \ifnum\LengthOfStack=\MaxItems % stack is full of items
  \Warn{WARNING: Stack is full...some items will be lost!}
 \else
  \Point=\StatusStack
  \advance\Point by \LengthOfStack
  \global\count\Point=\ItemSTATUS
  \Point=\NumStack
  \advance\Point by \LengthOfStack
  \global\count\Point=\ItemNUMBER
  \Point=\TypeStack
  \advance\Point by \LengthOfStack
  \global\count\Point=\ItemTYPE
  \Point\SpanStack
  \advance\Point by \LengthOfStack
  \global\count\Point=\ItemSPAN
  \Point=\BoxStack
  \advance\Point by \LengthOfStack
  \global\setbox\Point=\vbox{\copy\ItemBOX}
  \global\advance\LengthOfStack \@ne
  \ifnum\ItemTYPE=\Figure % used in \MakePage
   \global\MoreFigurestrue
  \else
   \global\MoreTablestrue
  \fi
 \fi
}

% item leaves stack
% #1=physical position of the item to be removed

\def\LeaveStack#1{%
 {\Iteration=#1
 \loop
 \ifnum\Iteration<\LengthOfStack
  \advance\Iteration \@ne
  \GetItemSTATUS{\Iteration}
   \advance\Point by \m@ne
   \global\count\Point=\ItemSTATUS
  \GetItemNUMBER{\Iteration}
   \advance\Point by \m@ne
   \global\count\Point=\ItemNUMBER
  \GetItemTYPE{\Iteration}
   \advance\Point by \m@ne
   \global\count\Point=\ItemTYPE
  \GetItemSPAN{\Iteration}
   \advance\Point by \m@ne
   \global\count\Point=\ItemSPAN
  \GetItemBOX{\Iteration}
   \advance\Point by \m@ne
   \global\setbox\Point=\vbox{\copy\ItemBOX}
 \repeat}
 \global\advance\LengthOfStack by \m@ne
}

% clean stack
% This routine scans through the stack and removes anything
% that does not have STATUS=\InStack.

\newif\ifStackNotClean

\def\CleanStack{%
 \StackNotCleantrue
 {\Iteration=\z@
  \loop
   \ifStackNotClean
    \GetItemSTATUS{\Iteration}
    \ifnum\ItemSTATUS=\InStack
     \advance\Iteration \@ne
     \else
      \LeaveStack{\Iteration}
    \fi
   \ifnum\LengthOfStack<\Iteration
    \StackNotCleanfalse
   \fi
 \repeat}
}

% Find item.
% This macro searches from the top to the bottom of the
% stack for an item of a specified type and number.
% #1=type, #2=number
% If the specified item is found, then \StackPointer is set
% to point to it, else \StackPointer=-1.
% This routine is used to find the next figure or table
% by number.

\def\FindItem#1#2{%
 \global\StackPointer=\m@ne % assume item isn't in stack for now
 {\Iteration=\z@
  \loop
  \ifnum\Iteration<\LengthOfStack
   \GetItemSTATUS{\Iteration}
   \ifnum\ItemSTATUS=\InStack
    \GetItemTYPE{\Iteration}
    \ifnum\ItemTYPE=#1
     \GetItemNUMBER{\Iteration}
     \ifnum\ItemNUMBER=#2
      \global\StackPointer=\Iteration
      \Iteration=\LengthOfStack % terminate loop
     \fi
    \fi
   \fi
  \advance\Iteration \@ne
 \repeat}
}

% Find next type
% This macro searches from the top to the bottom of the stack
% looking for the first item which has STATUS=\InStack.
% If it is a figure then a figure is what will be considered
% next by \MakePage else table.

\def\FindNext{%
 \global\StackPointer=\m@ne % assume stack is empty for now
 {\Iteration=\z@
  \loop
  \ifnum\Iteration<\LengthOfStack
   \GetItemSTATUS{\Iteration}
   \ifnum\ItemSTATUS=\InStack
    \GetItemTYPE{\Iteration}
   \ifnum\ItemTYPE=\Figure
    \ifMoreFigures
      \global\NextItem=\Figure
      \global\StackPointer=\Iteration
      \Iteration=\LengthOfStack % terminate loop
    \fi
   \fi
   \ifnum\ItemTYPE=\Table
    \ifMoreTables
      \global\NextItem=\Table
      \global\StackPointer=\Iteration
      \Iteration=\LengthOfStack % terminate loop
    \fi
   \fi
  \fi
  \advance\Iteration \@ne
 \repeat}
}

% Change status
% Macro to change the status of a specified item in stack.
% #1=item, #2=new status

\def\ChangeStatus#1#2{%
 \Point=\StatusStack
 \advance\Point by #1
 \global\count\Point=#2
}

%--------------------------------------------------------%
%                       MAKEPAGE                         %
%--------------------------------------------------------%

% This macro is called at the start of every new page
% including the first. It scans through the stack picking
% out items which should be placed on this page. It then
% leaves space for the items to be placed later. The routine
% terminates when either there is no room on the page to
% fit the next figure or table, or there are no more items
% in the stack.

\def\Zone{\InZoneA}

\ZoneBAdjust=\z@

\def\MakePage{% allocate space on this page for stack items
 \global\ZoneBSize=\PageHeight
 \global\TextSize=\ZoneBSize
 \global\ZoneAFullPagefalse
 \global\topskip=\TextLeading
 \MakePageInCompletetrue
 \MoreFigurestrue
 \MoreTablestrue
 \FigInZoneBfalse
 \FigInZoneCfalse
 \TabInZoneBfalse
 \TabInZoneCfalse
 \global\FirstSingleItemtrue
 \global\FirstZoneAtrue
 \global\setbox\ZoneABOX=\box\VOIDBOX
 \global\setbox\ZoneBBOX=\box\VOIDBOX
 \global\setbox\ZoneCBOX=\box\VOIDBOX
 \loop
  \ifMakePageInComplete
 \FindNext
 \ifnum\StackPointer=\m@ne
  \NextItem=\m@ne
  \MoreFiguresfalse
  \MoreTablesfalse
 \fi
 \ifnum\NextItem=\Figure
   \FindItem{\Figure}{\NextFigure}
   \ifnum\StackPointer=\m@ne \global\MoreFiguresfalse
   \else
    \GetItemSPAN{\StackPointer}
    \ifnum\ItemSPAN=\Single \def\Zone{\InZoneB}\relax
     \ifFigInZoneC \global\MoreFiguresfalse\fi
    \else
     \def\Zone{\InZoneA}
     \ifFigInZoneB \def\Zone{\InZoneC}\fi
    \fi
   \fi
   \ifMoreFigures\Print{}\FigureItems\fi
 \fi
\ifnum\NextItem=\Table
   \FindItem{\Table}{\NextTable}
   \ifnum\StackPointer=\m@ne \global\MoreTablesfalse
   \else
    \GetItemSPAN{\StackPointer}
    \ifnum\ItemSPAN=\Single\relax
     \ifTabInZoneC \global\MoreTablesfalse\fi
    \else
     \def\Zone{\InZoneA}
     \ifTabInZoneB \def\Zone{\InZoneC}\fi
    \fi
   \fi
   \ifMoreTables\Print{}\TableItems\fi
 \fi
   \MakePageInCompletefalse % assume page is complete
   \ifMoreFigures\MakePageInCompletetrue\fi
   \ifMoreTables\MakePageInCompletetrue\fi
 \repeat
%\Print{TextSize=\the\TextSize}
%\Print{ZoneBSize=\the\ZoneBSize}
 \ifZoneAFullPage
  \global\TextSize=\z@
  \global\ZoneBSize=\z@
  \global\vsize=\z@\relax
  \global\topskip=\z@\relax
  \vbox to \z@{\vss}
  \eject
 \else
 \global\advance\ZoneBSize by -\ZoneBAdjust
 \global\vsize=\ZoneBSize
 \global\hsize=\ColumnWidth
 \global\ZoneBAdjust=\z@
 \ifdim\TextSize<23pt
 \Warn{}
 \Warn{* Making column fall short: TextSize=\the\TextSize *}
 \vskip-\lastskip\eject\fi
 \fi
}

\def\MakeRightCol{% allocate space for the right column of text
 \global\TextSize=\ZoneBSize
 \MakePageInCompletetrue
 \MoreFigurestrue
 \MoreTablestrue
 \global\FirstSingleItemtrue
 \global\setbox\ZoneBBOX=\box\VOIDBOX
 \def\Zone{\InZoneB}
 \loop
  \ifMakePageInComplete
 \FindNext
 \ifnum\StackPointer=\m@ne
  \NextItem=\m@ne
  \MoreFiguresfalse
  \MoreTablesfalse
 \fi
 \ifnum\NextItem=\Figure
   \FindItem{\Figure}{\NextFigure}
   \ifnum\StackPointer=\m@ne \MoreFiguresfalse
   \else
    \GetItemSPAN{\StackPointer}
    \ifnum\ItemSPAN=\Double\relax
     \MoreFiguresfalse\fi
   \fi
   \ifMoreFigures\Print{}\FigureItems\fi
 \fi
 \ifnum\NextItem=\Table
   \FindItem{\Table}{\NextTable}
   \ifnum\StackPointer=\m@ne \MoreTablesfalse
   \else
    \GetItemSPAN{\StackPointer}
    \ifnum\ItemSPAN=\Double\relax
     \MoreTablesfalse\fi
   \fi
   \ifMoreTables\Print{}\TableItems\fi
 \fi
   \MakePageInCompletefalse % assume page is complete
   \ifMoreFigures\MakePageInCompletetrue\fi
   \ifMoreTables\MakePageInCompletetrue\fi
 \repeat
 \ifZoneAFullPage
  \global\TextSize=\z@
  \global\ZoneBSize=\z@
  \global\vsize=\z@\relax
  \global\topskip=\z@\relax
  \vbox to \z@{\vss}
  \eject
 \else
 \global\vsize=\ZoneBSize
 \global\hsize=\ColumnWidth
 \ifdim\TextSize<23pt
 \Warn{}
 \Warn{* Making column fall short: TextSize=\the\TextSize *}
 \vskip-\lastskip\eject\fi
\fi
}

\def\FigureItems{% Stack pointer points to next figure
 \Print{Considering...}
 \ShowItem{\StackPointer}
 \GetItemBOX{\StackPointer} % auto calculates ItemSIZE
 \GetItemSPAN{\StackPointer}
  \CheckFitInZone % check to see if item fits
  \ifnum\ItemFits=\Yes
   \ifnum\ItemSPAN=\Single
     \ChangeStatus{\StackPointer}{\InZoneB} % flag to be output
     \global\FigInZoneBtrue
     \ifFirstSingleItem
      \hbox{}\vskip-\BodgeHeight
     \global\advance\ItemSIZE by \TextLeading
     \fi
     \unvbox\ItemBOX\ItemSep
     \global\FirstSingleItemfalse
     \global\advance\TextSize by -\ItemSIZE% allocate space
     \global\advance\TextSize by -\TextLeading
   \else
    \ifFirstZoneA
     \global\advance\ItemSIZE by \TextLeading
     \global\FirstZoneAfalse\fi
    \global\advance\TextSize by -\ItemSIZE
    \global\advance\TextSize by -\TextLeading
    \global\advance\ZoneBSize by -\ItemSIZE
    \global\advance\ZoneBSize by -\TextLeading
    \ifFigInZoneB\relax
     \else
     \ifdim\TextSize<3\TextLeading
     \global\ZoneAFullPagetrue
     \fi
    \fi
    \ChangeStatus{\StackPointer}{\Zone}
    \ifnum\Zone=\InZoneC \global\FigInZoneCtrue\fi
  \fi
   \Print{TextSize=\the\TextSize}
   \Print{ZoneBSize=\the\ZoneBSize}
  \global\advance\NextFigure \@ne
   \Print{This figure has been placed.}
  \else
   \Print{No space available for this figure...holding over.}
   \Print{}
   \global\MoreFiguresfalse
  \fi
}

\def\TableItems{% Stack pointer points to next table
 \Print{Considering...}
 \ShowItem{\StackPointer}
 \GetItemBOX{\StackPointer} % auto calculates ItemSIZE
 \GetItemSPAN{\StackPointer}
  \CheckFitInZone % check to see of item fits in Zone
  \ifnum\ItemFits=\Yes
   \ifnum\ItemSPAN=\Single
    \ChangeStatus{\StackPointer}{\InZoneB}
     \global\TabInZoneBtrue
     \ifFirstSingleItem
      \hbox{}\vskip-\BodgeHeight
     \global\advance\ItemSIZE by \TextLeading
     \fi
     \unvbox\ItemBOX\ItemSep
     \global\FirstSingleItemfalse
     \global\advance\TextSize by -\ItemSIZE
     \global\advance\TextSize by -\TextLeading
   \else
    \ifFirstZoneA
    \global\advance\ItemSIZE by \TextLeading
    \global\FirstZoneAfalse\fi
    \global\advance\TextSize by -\ItemSIZE
    \global\advance\TextSize by -\TextLeading
    \global\advance\ZoneBSize by -\ItemSIZE
    \global\advance\ZoneBSize by -\TextLeading
    \ifFigInZoneB\relax
     \else
     \ifdim\TextSize<3\TextLeading
     \global\ZoneAFullPagetrue
     \fi
    \fi
    \ChangeStatus{\StackPointer}{\Zone}
    \ifnum\Zone=\InZoneC \global\TabInZoneCtrue\fi
   \fi
%   \Print{TextSize=\the\TextSize}
%   \Print{ZoneBSize=\the\ZoneBSize}
  \global\advance\NextTable \@ne
   \Print{This table has been placed.}
  \else
  \Print{No space available for this table...holding over.}
   \Print{}
   \global\MoreTablesfalse
  \fi
}

% These macros check to see if an item of ItemSIZE will
% fit in a particular zone. If it will, then ItemFits
% will be set true else false.

\def\CheckFitInZone{%
{\advance\TextSize by -\ItemSIZE
 \advance\TextSize by -\TextLeading
 \ifFirstSingleItem
  \advance\TextSize by \TextLeading
 \fi
 \ifnum\Zone=\InZoneA\relax
  \else \advance\TextSize by -\ZoneBAdjust
 \fi
 \ifdim\TextSize<3\TextLeading \global\ItemFits=\No
 \else \global\ItemFits=\Yes\fi}
}

\def\BeginOpening{%
  % start 9pt a.s.a.p. so that \New.. commands get a chance to init.
  \ninepoint
  \thispagestyle{titlepage}%
  \global\setbox\ItemBOX=\vbox\bgroup%
    \hsize=\PageWidth%
    \hrule height \z@
    \ifsinglecol\vskip 6pt\fi % Bodge, to get same pos. as two-column!
}

\let\begintopmatter=\BeginOpening  %  alias for \BeginOpening

\def\EndOpening{%
  \One%  1 line fixed space below opening
  \egroup
  \ifsinglecol
    \box\ItemBOX%
    \vskip\TextLeading plus 2\TextLeading% var. space: min 1, max 3 lines
    \@noafterindent
  \else
    \ItemNUMBER=\z@%
    \ItemTYPE=\Figure
    \ItemSPAN=\Double
    \ItemSTATUS=\InStack
    \JoinStack
  \fi
}

% Figures

\newif\if@here  \@herefalse

\def\no@float{\global\@heretrue}
\let\nofloat=\relax % only enabled for one column

\def\beginfigure{%
  \@ifstar{\global\@dfloattrue \@bfigure}{\global\@dfloatfalse \@bfigure}%
}

\def\@bfigure#1{%
  \par
  \if@dfloat
    \ItemSPAN=\Double
    \TEMPDIMEN=\PageWidth
  \else
    \ItemSPAN=\Single
    \TEMPDIMEN=\ColumnWidth
  \fi
  \ifsinglecol
    \TEMPDIMEN=\PageWidth
  \else
    \ItemSTATUS=\InStack
    \ItemNUMBER=#1%
    \ItemTYPE=\Figure
  \fi
  \bgroup
    \hsize=\TEMPDIMEN
    \global\setbox\ItemBOX=\vbox\bgroup
      \eightpoint\nostb@ls{10pt}%
      \let\caption=\fig@caption
      \ifsinglecol \let\nofloat=\no@float\fi
}

\def\fig@caption#1{%
  \vskip 5.5pt plus 6pt%
  \bgroup % grouping and size change needed for plate pages
    \eightpoint\nostb@ls{10pt}%
    \setbox\TEMPBOX=\hbox{#1}%
    \ifdim\wd\TEMPBOX>\TEMPDIMEN
      \noindent \unhbox\TEMPBOX\par
    \else
      \hbox to \hsize{\hfil\unhbox\TEMPBOX\hfil}%
    \fi
  \egroup
}

\def\endfigure{%
  \par\egroup % end \vbox
  \egroup
  \ifsinglecol
    \if@here \midinsert\global\@herefalse\else \topinsert\fi
      \unvbox\ItemBOX
    \endinsert
  \else
    \JoinStack
    \Print{Processing source for figure \the\ItemNUMBER}%
  \fi
}

% Tables

\newbox\tab@cap@box
\def\tab@caption#1{\global\setbox\tab@cap@box=\hbox{#1\par}}

\newtoks\tab@txt@toks
\long\def\tab@txt#1{\global\tab@txt@toks={#1}\global\table@txttrue}

\newif\iftable@txt  \table@txtfalse
\newif\if@dfloat    \@dfloatfalse

\def\begintable{%
  \@ifstar{\global\@dfloattrue \@btable}{\global\@dfloatfalse \@btable}%
}

\def\@btable#1{%
  \par
  \if@dfloat
    \ItemSPAN=\Double
    \TEMPDIMEN=\PageWidth
  \else
    \ItemSPAN=\Single
    \TEMPDIMEN=\ColumnWidth
  \fi
  \ifsinglecol
    \TEMPDIMEN=\PageWidth
  \else
    \ItemSTATUS=\InStack
    \ItemNUMBER=#1%
    \ItemTYPE=\Table
  \fi
  \bgroup
    \eightpoint\nostb@ls{10pt}%
    \global\setbox\ItemBOX=\vbox\bgroup
      \let\caption=\tab@caption
      \let\tabletext=\tab@txt
      \ifsinglecol \let\nofloat=\no@float\fi
}

\def\endtable{%
  \par\egroup % end \vbox
  \egroup
  \setbox\TEMPBOX=\hbox to \TEMPDIMEN{%
    \eightpoint\nostb@ls{10pt}%
    \hss
    \vbox{%
      \hsize=\wd\ItemBOX
      \ifvoid\tab@cap@box
      \else
        \noindent\unhbox\tab@cap@box
        \vskip 5.5pt plus 6pt%
      \fi
      \box\ItemBOX
      \iftable@txt
        \vskip 10pt%
        \noindent\the\tab@txt@toks
        \global\table@txtfalse
      \fi
    }%
    \hss
  }%
  \ifsinglecol
    \if@here \midinsert\global\@herefalse\else \topinsert\fi
      \box\TEMPBOX
    \endinsert
  \else
    \global\setbox\ItemBOX=\box\TEMPBOX
    \JoinStack
    \Print{Processing source for table \the\ItemNUMBER}%
  \fi
}

\def\UnloadZoneA{%
\FirstZoneAtrue
 \Iteration=\z@
  \loop
   \ifnum\Iteration<\LengthOfStack
    \GetItemSTATUS{\Iteration}
    \ifnum\ItemSTATUS=\InZoneA
     \GetItemBOX{\Iteration}
     \ifFirstZoneA \vbox to \BodgeHeight{\vfil}%
     \FirstZoneAfalse\fi
     \unvbox\ItemBOX\ItemSep
     \LeaveStack{\Iteration}
     \else
     \advance\Iteration \@ne
   \fi
 \repeat
}

\def\UnloadZoneC{%
\Iteration=\z@
  \loop
   \ifnum\Iteration<\LengthOfStack
    \GetItemSTATUS{\Iteration}
    \ifnum\ItemSTATUS=\InZoneC
     \GetItemBOX{\Iteration}
     \ItemSep\unvbox\ItemBOX
     \LeaveStack{\Iteration}
     \else
     \advance\Iteration \@ne
   \fi
 \repeat
}

%--------------------------------------------------------%
%                         DIAGNOSTICS                    %
%--------------------------------------------------------%

\def\ShowItem#1{% Show details of on item entry in stack
  {\GetItemAll{#1}
  \Print{\the#1:
  {TYPE=\ifnum\ItemTYPE=\Figure Figure\else Table\fi}
  {NUMBER=\the\ItemNUMBER}
  {SPAN=\ifnum\ItemSPAN=\Single Single\else Double\fi}
  {SIZE=\the\ItemSIZE}}}
}

\def\ShowStack{% 
 \Print{}
 \Print{LengthOfStack = \the\LengthOfStack}
 \ifnum\LengthOfStack=\z@ \Print{Stack is empty}\fi
 \Iteration=\z@
 \loop
 \ifnum\Iteration<\LengthOfStack
  \ShowItem{\Iteration}
  \advance\Iteration \@ne
 \repeat
}

\def\B#1#2{%
\hbox{\vrule\kern-0.4pt\vbox to #2{%
\hrule width #1\vfill\hrule}\kern-0.4pt\vrule}
}

%-------------------------------------------------------%
%             SINGLE COLUMN OUTPUT ROUTINE              %
%-------------------------------------------------------%

\newif\ifsinglecol   \singlecolfalse

\def\onecolumn{%
  \global\output={\singlecoloutput}%
  \global\hsize=\PageWidth
  \global\vsize=\PageHeight
  \global\ColumnWidth=\hsize
  \global\TextLeading=12pt
  \global\Leading=12
  \global\singlecoltrue
  \global\let\onecolumn=\relax%         stop them using \onecolumn again
  \global\let\footnote=\sing@footnote%  enable footnotes
  \global\let\vfootnote=\sing@vfootnote
  \ninepoint % reset \baselineskip after leading change
  \message{(Single column)}%
}

\def\singlecoloutput{%
  \shipout\vbox{\PageHead\vbox to \PageHeight{\pagebody\vss}\PageFoot}%
  \advancepageno
  \ifplate@page
    \shipout\vbox{%
      \sp@pagetrue
      \def\sp@type{plate}%
      \global\plate@pagefalse
      \PageHead\vbox to \PageHeight{\unvbox\plt@box\vfil}\PageFoot%
    }%
    \message{[plate]}%
    \advancepageno
  \fi
  \ifnum\outputpenalty>-\@MM \else\dosupereject\fi%
}

\def\ItemSep{\vskip\ItemSepamount\relax}

\def\ItemSepbreak{\par\ifdim\lastskip<\ItemSepamount
  \removelastskip\penalty-200\ItemSep\fi%
}

% Modify plain's \endinsert so that the mn's spacing is used

\let\@@endinsert=\endinsert % save plain's original \endinsert

\def\endinsert{\egroup % finish the \vbox
  \if@mid \dimen@\ht\z@ \advance\dimen@\dp\z@ \advance\dimen@12\p@
    \advance\dimen@\pagetotal \advance\dimen@-\pageshrink
    \ifdim\dimen@>\pagegoal\@midfalse\p@gefalse\fi\fi
  \if@mid \ItemSep\box\z@\ItemSepbreak
  \else\insert\topins{\penalty100 % floating insertion
    \splittopskip\z@skip
    \splitmaxdepth\maxdimen \floatingpenalty\z@
    \ifp@ge \dimen@\dp\z@
    \vbox to\vsize{\unvbox\z@\kern-\dimen@}% depth is zero
    \else \box\z@\nobreak\ItemSep\fi}\fi\endgroup%
}

% Footnotes (only enabled in single column)

\def\gobbleone#1{}
\def\gobbletwo#1#2{}
\let\footnote=\gobbletwo % Gobble footnote's unless enabled by \onecolumn
\let\vfootnote=\gobbleone

\def\sing@footnote#1{\let\@sf\empty % parameter #2 (the text) is read later
  \ifhmode\edef\@sf{\spacefactor\the\spacefactor}\/\fi
  \hbox{$^{\hbox{\eightpoint #1}}$}\@sf\sing@vfootnote{#1}%
}

\def\sing@vfootnote#1{\insert\footins\bgroup\eightpoint\b@ls{9pt}%
  \interlinepenalty\interfootnotelinepenalty
  \splittopskip\ht\strutbox % top baseline for broken footnotes
  \splitmaxdepth\dp\strutbox \floatingpenalty\@MM
  \leftskip\z@skip \rightskip\z@skip \spaceskip\z@skip \xspaceskip\z@skip
  \noindent $^{\scriptstyle\hbox{#1}}$\hskip 4pt%
    \footstrut\futurelet\next\fo@t%
}

% Kill footnote rule
\def\footnoterule{\kern-3\p@ \hrule height \z@ \kern 3\p@}

\skip\footins=19.5pt plus 12pt minus 1pt
\count\footins=1000
\dimen\footins=\maxdimen

% for footnotes in double column: use \note{$\star$}{footnote}
\def\note#1#2{%
  \let\@sf=\empty \ifhmode\edef\@sf{\spacefactor\the\spacefactor}\/\fi
  #1\insert\footins\bgroup
    \eightpoint\b@ls{10pt}\rm
    \interlinepenalty\interfootnotelinepenalty
%    \splittopskip\ht\strutbox % top baseline for broken footnotes
    \splitmaxdepth\dp\strutbox \floatingpenalty\@MM
    \leftskip\z@skip \rightskip\z@skip \spaceskip\z@skip \xspaceskip\z@skip
    \noindent\footstrut #1$\,$#2\strut\par
  \egroup
  \@sf\relax}

% Landscape

\def\landscape{%
  \global\TEMPDIMEN=\PageWidth
  \global\PageWidth=\PageHeight
  \global\PageHeight=\TEMPDIMEN
  \global\let\landscape=\relax%         stop them using \landscape again.
  \onecolumn
  \message{(landscape)}%
  \raggedbottom
}

%-------------------------------------------------------%
%               TWO COLUMN OUTPUT ROUTINE               %
%-------------------------------------------------------%

% Very slight redefinition of the \output routine of mn.tex, to allow footnotes.
\output{%
  \ifLeftCOL
    \global\setbox\LeftBOX=\vbox to \ZoneBSize{\box255\unvbox\ZoneBBOX
      \ifvoid\footins\else
        \vskip\skip\footins\unvbox\footins\fi
    }%
    \global\LeftCOLfalse
    \MakeRightCol
  \else
    \setbox\RightBOX=\vbox to \ZoneBSize{\box255\unvbox\ZoneBBOX
      \ifvoid\footins\else
        \vskip\skip\footins\unvbox\footins\fi
    }%
    \setbox\MidBOX=\hbox{\box\LeftBOX\hskip\ColumnGap\box\RightBOX}%
    \setbox\PageBOX=\vbox to \PageHeight{%
      \UnloadZoneA\box\MidBOX\UnloadZoneC}%
    \shipout\vbox{\PageHead\vbox to \PageHeight{\box\PageBOX\vss}\PageFoot}%
    \advancepageno
    \ifplate@page
      \shipout\vbox{%
        \sp@pagetrue
        \def\sp@type{plate}%
        \global\plate@pagefalse
        \PageHead\vbox to \PageHeight{\unvbox\plt@box\vfil}\PageFoot%
      }%
      \message{[plate]}%
      \advancepageno
    \fi
    \global\LeftCOLtrue
    \CleanStack
    \MakePage
  \fi
}

% Startup message

\Warn{\start@mess}

\newif\ifCUPmtplainloaded % for use in documents
\ifprod@font
  \global\CUPmtplainloadedtrue
\fi

\def\mnmacrosloaded{} % so articles can see if a format file has been used.

\catcode `\@=12 % @ signs are non-letters

% \dump

% end of mn.tex

\fi
\input epsf
\epsfverbosetrue
%\pageoffset{-2.5pc}{0pc}

%\Referee

\begintopmatter

\title{Australia Telescope Search for Cosmic Microwave 
       Background Anisotropy}

\author{Ravi~Subrahmanyan$^{1,2}$,
         Michael~J.~Kesteven$^2$,
         Ronald~D.~Ekers$^2$,
         Malcolm~Sinclair$^2$,
         Joseph~Silk$^3$}

\affiliation{$^1$ Raman Research Institute, Sadashivanagar, 
             Bangalore 560 080, India}
\vskip 0.1 truecm
\affiliation{$^2$ Australia Telescope National Facility, CSIRO, PO Box 76,
             Epping, NSW 2121, Australia}
\vskip 0.1 truecm
\affiliation{$^3$ Departments of Astronomy and Physics and Centre for Particle
             Astrophysics, University of California, Berkeley, CA 94720, USA}

\shortauthor{R.~Subrahmanyan et al.}

\shorttitle{ATCA observations for CMB anisotropy}

\abstract {
In an attempt to detect cosmic microwave background (CMB) anisotropy on
arcmin scales, 
we have made an 8.7-GHz image of a sky region with a 
resolution of two arcmin and
high surface brightness sensitivity using the Australia Telescope Compact
Array (ATCA) in an ultra-compact configuration.  
The foreground discrete-source confusion was
estimated from observations with 
higher resolution at the same frequency and in a scaled array at a lower 
frequency.  Following the subtraction of the foreground confusion,
the field shows no features in excess of the instrument noise. This limits
the CMB anisotropy flat-band power 
to $Q_{flat} < 23.6~\mu$K with 95 per cent confidence; the 
ATCA filter function$^{*}$
$F_{l}$ in multipole $l$-space peaks at $l_{eff}=4700$ and has 
half maximum values at
$l$ = 3350 and 6050.  
	  }

\keywords {cosmic microwave background
           -- cosmology: observations
           -- techniques: interferometric.}

\maketitle

\section{Introduction}

Anisotropies in the cosmic microwave background (CMB)
\note{}{$^{*}$The ATCA filter function is available at the website} 
\note{}{{\it ~~www.atnf.csiro.au/research/cmbr/cmbr\_atca.html}}
are usually described
in terms of the coefficients $C_{lm}$ of their spherical harmonic
decomposition; $C_{l} = \langle \mid C_{lm} \mid ^{2} \rangle$
represents the anisotropy power at multipole order $l$.  Sachs-Wolfe
anisotropies in a scale-invariant matter power spectrum give flat
$l(l+1)C_{l}$ and any experiment sensitive to CMB anisotropies in an $l$-space
window $F_{l}$ may be expected to measure a sky temperature variance
$$ 
 (\Delta T)^{2} = \sum_{l} {{6(2l+1)}\over{5l(l+1)}} 
 (Q_{flat})^{2} F_{l}, \eqno\stepeq
$$
\noindent where $Q_{flat}$ denotes the quadrupole normalization of a 
flat CMB anisotropy spectrum.  
The observing scheme and data analysis procedures determine the window
function $F_{l}$ for any experiment.  The implications of any observed 
CMB temperature variance $(\Delta T)^{2}$ may be quoted
in fairly model independent terms by the inferred $Q_{flat}$.  This
$Q_{flat}$ may be viewed as being a measure of the CMB anisotropy
power spectral density within the $l$-space window defined by $F_{l}$.

Recent observations of the anisotropy in the CMB on large and intermediate 
angular scales ($l \la 500$) have begun to provide interesting
constraints on theories of structure formation and the parameters of
cosmological models (see, for example, Lineweaver and Barbosa (1997) for
an updated compilation of data points and a discussion of their implications).

Primary CMB anisotropies at small angular scales 
($l \ga 500$) are expected to be relatively
damped in most structure formation models owing to the the thickness of the
last scattering surface and the diffusion damping of sub-horizon scale
baryon fluctuations in the pre-recombination epoch.
However, flat-band powers comparable to the $Q_{flat}$ detected
by {\it COBE DMR} (Bennett et al. 1996) may be predicted at multipoles 
$l \ga 500$ in certain cosmological-constant ($\Lambda$)
and open-universe manifestations of 
baryon isocurvature models (Hu \& Sugiyama 1994; Hu, Bunn \& Sugiyama 1995).  
The anisotropy
power at large $l$ may be critically dependent on the reionization
history: small-scale anisotropies may be suppressed by early reionization.

Secondary anisotropies may be generated at the later last-scattering-surface
in a reionized universe
predominantly owing to second-order mode coupling between density
perturbations and bulk velocities (the `Vishniac effect'); this could
significantly contribute to arcmin-scale ($l \ga 10^{3}$) anisotropies
particularly if the ionization fraction is high at late times (Hu, Scott
\& Silk 1994).  Persi et al. (1995) estimate that this second-order
Doppler effect may contribute a band power $Q_{flat} \approx 0.2~\mu$K
at $l \approx 10^{3}$ in CDM and CDM+$\Lambda$ universes and that the
anisotropy power may be an order of magnitude higher in certain 
baryon isocurvature models.

Jones et al. (1997) report the detection of a $\sim 100~\mu$K negative feature
in a 2-arcmin resolution image of a sky patch that has no obvious cluster 
of galaxies along the line of sight in either optical 
or {\it ROSAT} X-ray images.  A sensitive low-resolution image
of a `blank' field with the VLA (Richards et al. 1997) is also reported
to show a negative feature approximately 
$25^{\prime\prime} \times 65^{\prime\prime}$ in size and 
with a peak central decrement of $-250~\mu$K.  It has been suggested 
that these may be CMB decrements owing to the 
inverse-Compton scattering of CMB photons, the 
Sunyaev-Zeldovich (S-Z) effect, in distant concentrations of hot gas.
It may be noted that Hattori et al. (1997) recently reported the 
discovery, in its X-ray emission, of a hot gas concentration at a redshift 
$z \approx 1$  whose properties are similar to the gaseous halos in rich
clusters of galaxies; however, the `cluster' appears to have only one 
visible galaxy.   

Anisotropies arising from the S-Z effect,
in a cosmological population of groups and clusters of galaxies containing hot
intra-cluster gas  may be an important
cause of anisotropy power at $l \ga 10^{3}$ (Persi et al. 1995; Bond \& Myers
1996).  The S-Z anisotropies are generically
non-Gaussian and are expected to have phase-correlations between different
$l$ modes; their statistical description requires higher-order correlations 
apart from the $C_{l}$ power spectra.
Hot gas in groups and clusters in scale-invariant 
CDM universes normalized to give $\sigma_{8} = 1$ (the magnitude of dynamical
clustering at the present time is quantified by $\sigma_{8}$ which is 
the rms mass fluctuations in $8 h^{-1}$-Mpc spheres) is expected to
contribute band powers $Q_{flat} \approx 2~\mu$K 
at $l \approx 1$--$5 \times 10^{3}$
(Bond \& Myers 1996).  In {\it COBE}-normalized tilted CDM, MDM and models in 
which the shape factor $\Gamma = \Omega_{\circ} h$ is approximately 0.2, 
the predicted $\sigma_{8}$
agrees better with galaxy clustering and in these models the band powers at
large $l$ are 1--2 orders of magnitude smaller. The dominant contribution 
at high $l$ may come from quasar-ionized hot gas 
bubbles (Aghanim et al. 1996), whose existence is based on 
plausible theoretical inferences.

Computations of the expected secondary anisotropies involve non-linear
gravitational dynamics and hydrodynamic
simulations and are critically
dependent on the thermal history of the gas that may in turn depend on the 
astrophysical evolution in populations that cause the ionization.
It follows that
observations of the arcmin-scale anisotropy, corresponding to measurements
of the CMB anisotropy at $l \ga 10^{3}$, could constrain the structure
formation theories.

\section{Observations with the Australia Telescope}

The advantages of Fourier-Synthesis imaging telescopes, and in
particular the design features of the Australia Telescope
Compact Array (ATCA; see The Australia Telescope 1992) 
that make it specifically
advantageous for high-brightness-sensitivity imaging,  were detailed
in Subrahmanyan et al. (1993) and we restrict ourselves to giving a
synopsis of the methodology here. 
Our observing strategy has been to make full Earth-rotation synthesis
observations of `empty' fields in a special ultra-compact 122-m array
configuration -- with five 22-m diameter antennae located 30.6~m apart
in an E-W line -- in order to maximize the brightness sensitivity of the
imaging.  The three baselines between antennae spaced 61~m apart are 
used to construct a model of the foreground confusion; this is then
subtracted from all of the visibility data and the four baselines
between antennae 30.6~m apart are used to synthesize an image with
high brightness sensitivity.  
Because the confusing sources are measured simultaneously and using baselines
between the very same antennae, errors in the estimation of confusion
owing to variability in the foreground sources
and calibration errors are eliminated.  Deconvolution errors are avoided
because the confusion is estimated as a model fit to visibility data.
These observations were made at the highest available frequency
of 8.7~GHz to minimize discrete source confusion.
  
The field was separately observed in a nearly scaled 244-m array
at 4.7~GHz, with five ATCA antennae spaced 61-m apart along the E-W
line, to  examine the spectral indices of any features identified in the
region as foreground sources and ensure that the sources subtracted
indeed have spectra consistent with optically-thin thermal or
synchrotron emission.  If the 4.7-GHz images are made with the same
sensitivity to point-source flux density as the 8.7-GHz images, the
lower-frequency images that are made with the same angular resolution
would be less sensitive to CMB temperature
fluctuations by a factor $(4.7/8.7)^{\alpha}$, where $\alpha = 2$ is 
the spectral index of the CMB anisotropy at these frequencies. 
However, the 4.7-GHz images would have a relatively greater sensitivity
to extended synchrotron emission by a factor $(4.7/8.7)^{\alpha}$, where
$\alpha \la -0.7$ is now the steep spectral index of extended synchrotron
sources.  To summarize, the 4.7-GHz image could potentially
reveal extended foreground sources that may be
resolved, and therefore absent, in the 61-m baseline data obtained at
8.7~GHz.

Sky regions -- selected using radio source catalogues to be relatively 
devoid of bright sources -- were first imaged at 20-cm wavelength with the
ATCA.  An examination of these images led to the selection of
a field centre which had no sources with flux densities 
exceeding 1~mJy
(at 20-cm wavelength) within 7~arcmin radius (the ATCA antennae have
the first null of the 8.7-GHz primary beam at 7~arcmin).  The J2000.0-epoch
coordinates of the field centre are 
RA: 03$^{h}$~16$^{m}$~26\fs00, DEC: $-$49\degr~47\arcmin~57\farcs00.
The field has been chosen to be located close to declination $-50^{\circ}$
so that the projected antenna spacing is
as close as possible without shadowing to the 22-m antenna diameter 
at large hour angles: this maximizes the brightness sensitivity of the array.  

The field was originally observed in 1991 July-August and December in the 
122-m array and using a pair of 128-MHz bands centred at 8640 and 8768~MHz; 
results of the observations made during these periods were
reported in Subrahmanyan et al. (1993). Subsequently, the 8.7-GHz
front-end amplifiers were changed to HEMT devices and the system temperature
at this frequency improved from about 75 to 43~K.  The field was
reobserved in 1994 July in the 122-m array using the same pair of bands.
The total effective observing time obtained in this array 
now corresponds to about 50~h 
integration with a system temperature of 43~K and two 128-MHz bands.
The rms sensitivity of the imaging has improved by a factor 1.6 as a
result of the additional observations made with the improved receivers.
All observations were made in dual polarizations.
The array phase centres during the
observations were offset about $1^{\circ}$
from the antenna pointing centres so that 
imaging artefacts that often appear at the phase centre may be well
removed from the sky region of interest.  Off-line,
the visibility data were phase-corrected to align the array phase centres
with the field centre positions, calibrated in amplitude, phase and for the
band-pass response, averaged in frequency over a useful band of 112~MHz.  
The flux-density scale was set by adopting values
of 2.84 and 2.79~Jy respectively for the primary calibrator source 
PKS~1934$-$638 at the frequencies 8640 and 8768~MHz (Reynolds 1994).  

In 1995 March, the field was observed in a 244-m array at 4800~MHz.
The observing and calibration procedure adopted was the same as for the
122-m array observations; the flux-density scale was set by adopting a
value of 6.22~Jy for the primary calibrator PKS~1934$-$638 at this frequency.  

All images shown below were made using visibilities 
with `natural' weighting
so as to obtain a high signal-to-noise ratio, the
gridding, Fourier transformations and deconvolution, if attempted, were done 
using the {\sc aips} routine {\sc imagr}.  All displayed images are centred at the
coordinates of the field centre and, unless explicitely stated, have not been
deconvolved.  All images shown are in Stokes I and have not been corrected
for the attenuation owing to the primary beam. The locus of the first null
in the primary beam pattern at 8704~MHz, located at a radius of 7~arcmin,
is shown in all the images as a dot-dashed circle.

\section{Analysis of the observed field}

\beginfigure{1}
  \nofloat
  \epsfbox[45 35 275 275]{fig1.ps}
  \caption{{\bf Figure 1.} Deconvolved image of the field at 4800~MHz made with
   a beam of $96 \times 72$ arcsec$^{2}$ at a position angle (p.a.) 
   of $52^{\circ}$.  Contours at 27 $\mu$Jy~beam$^{-1}~\times~(-$3, 3,
   4, 6, 8, 12, 16, 24, 32).  The half-maximum size of the synthesized beam 
   is shown in the bottom right corner as a filled ellipse.  In this figure, as
   also the following three figures, the locus of the first null in the primary 
   beam pattern at 8704~MHz is shown as a dot-dashed circle.}
\endfigure

\beginfigure{2}
  \nofloat
  \epsfbox[45 35 275 275]{fig2.ps}
  \caption{{\bf Figure 2.} An image of the field at 8704~MHz made using
   just the 30.6-m baselines.  The resolution is about 2.2 arcmin.
   Contours at 22 $\mu$Jy~beam$^{-1}~\times~(-$10, 
   $-$8, $-$6, $-$4, 4, 6, 8, 10, 12, 14, 16).}
\endfigure

\beginfigure{3}
  \nofloat
  \epsfbox[45 35 275 275]{fig3.ps}
  \caption{{\bf Figure 3.} A deconvolved image of the field at 8704~MHz made
   using just the 61-m baselines.  The image has a beam FWHM of 
   $71 \times 57$ arcsec$^{2}$ at a position angle (p.a.) 
   of $0^{\circ}$. Contours at 25 $\mu$Jy~beam$^{-1}~\times~(-$4, $-$3, 
   $-$2, 2, 3, 4, 6, 8, 10). The half-maximum size of the synthesized beam 
   is shown in the bottom right corner as a filled ellipse.}
\endfigure

\beginfigure{4}
  \nofloat
  \epsfbox[45 35 275 275]{fig4.ps}
  \caption{{\bf Figure 4.} The residual image of the field at 8704~MHz 
   following the subtraction of the confusion model; this image has been 
   made using just the 30.6-m baselines.  The resolution is about 2.2 arcmin.
   Contours at 22 $\mu$Jy~beam$^{-1}~\times~(-$3, $-$2, $-$1, 1, 2, 3).}
\endfigure

An image of the field made at 4.8~GHz using the 61 and 122-m baselines of the
244-m array observations is shown in
Fig.~1.  The image has been deconvolved and has a resolution of
1.4~arcmin.  Within a radius of 3~arcmin, corresponding to the half-power
radius of the ATCA primary beam at 8.7~GHz, three prominent sources are
apparent and all these are detected with peak flux density exceeding 10
times the image thermal noise.

In Fig.~2 we show an image of this field at 8.7~GHz that has been constructed 
using just the four 30.6-m baselines.  This image has not been deconvolved, but
clearly shows three peaks at the positions of the three sources apparent in
the 4.8-GHz image.  It may be noted that the synthesized beam sidelobes
are very large for this image because of the poor $u,v$ coverage (only a
single spacing -- 30.6~m -- has been used in the imaging).

The low-resolution 8.7-GHz image of the field (Fig.~2) has a high surface
brightness sensitivity, but is confusion limited owing to discrete sources
in the field.  We estimate the foreground confusion owing to discrete
sources in this field using the data obtained in the longer ($> 30.6$~m)
baselines.  However, we have restricted ourselves to using just the 61-m
baseline (omitting the 92 and 122~m baselines) for deriving the confusion model
so that the synthesized beam is only a factor of two smaller as compared to
that for the high-surface-brightness image (Fig.~2) from which the
cofusion is to be subtracted.  This may ensure that confusion structures
on scales up to about an arcmin, if present, 
may be included in the confusion model.

We next show a deconvolved image of the field made at 8.7~GHz using the
61-m baselines (Fig.~3).  Three sources appear once again, we estimate their
positions and flux densities at 8.7~GHz from this deconvolved image and
use these parameters as an input model while fitting a 3-component
model to the 61-m visibility data.  The fit estimated the flux densities
of the components to be 274, 182 and 110 $\mu$Jy.  The primary beam at
8.7~GHz is expected to have attenuated the source intensities by
factors 0.57, 0.39 and 0.51 respectively.  Correcting for this
attenuation, the components ought to
have flux densities  481, 467 and 216 $\mu$Jy respectively.       
We have fit the 4.8-GHz visibilities measured using the 122-m baselines
(equivalent in angular resolution to the 61~m baselines at 8.7~GHz) 
to a 3-component model that has component positions fixed at the locations
of the sources in the 8.7-GHz, 61-m visibilities: 
the flux densities of these components are estimated to be 790, 335 and 
262~$\mu$Jy 
respectively.  Correcting for the attenuation owing to the primary beam
at 4.8~GHz, we estimate the true flux densities of these components
to be 934, 439 and 319~$\mu$Jy respectively.  The derived spectral
indices of the components are $-1.1$, 0.1 and $-0.7$ (we define the spectral
index $\alpha$ as $S_{\nu} \sim \nu^{\alpha}$); the components do
not have spectral indices $\alpha = 2$ that would be expected of
CMB anisotropies at these frequencies.  

The three-component model for the foreground confusion, derived from the
fit to the 60-m visibilities obtained at 8.7~GHz, was subtracted from
all of the 8.7-GHz visibility data.
Following subtraction of the confusion, we constructed an image of the field 
at 8.7~GHz using just the 30.6-m baselines (Fig.~4).  No residual features
are apparent in this image.
We weighted the image pixel intensities by the primary beam pattern
and determined the weighted rms in the primary beam region of
the field to be $21.2~\mu$Jy~beam$^{-1}$. 
The image pixel variance appears consistent
with that expected from the telescope thermal noise (see section~4.1), 
and in sections~5 and 6 below we use this
residual image to derive limits on random-phase CMB anisotropy on the
2-arcmin scale corresponding to the resolution in this image.

\section{Contributors to the image variance}

\beginfigure*{5}
  \nofloat
  \epsfbox[-100 -25 422 226]{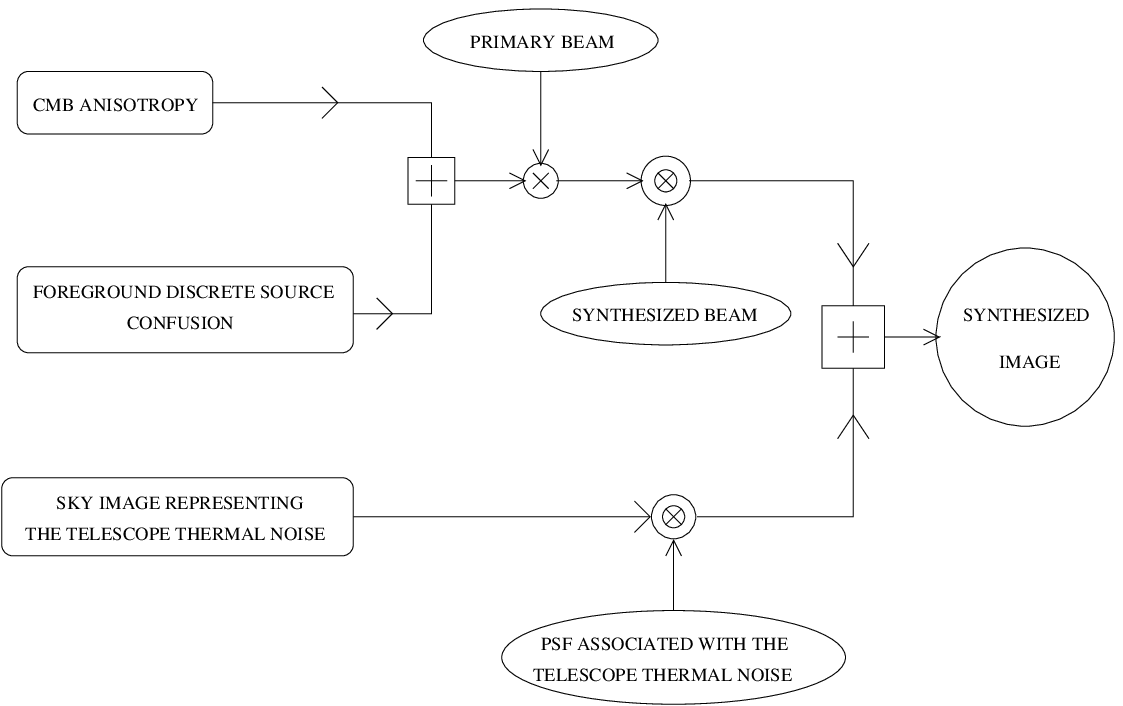}
  \caption{{\bf Figure 5.} The signal flow path.  It may be noted that the
  point spread function (PSF) corresponding to the sky 
  sources -- the synthesized
  beam -- may differ from the PSF associated with the telescope thermal noise.}
\endfigure

The processes which lead to the synthesized image are shown
schematically in Fig.~5.  The telescope
thermal noise, foreground discrete sources and possibly CMB anisotropy
contribute to the variance in the synthesized image.  CMB anisotropy
and foreground radio sources on the sky are attenuated by the primary-beam
pattern of the antennae and convolved by the synthesized beam; their
contribution may equivalently be viewed as a visibility-domain 
spatial-frequency filtering of the sky
signals as viewed through the primary beam.  The telescope thermal noise may
be considered to be additive Gaussian random fluctuations, that are 
characterized by an appropriate angular coherence function, but whose
distribution on the sky image has no dependence on the primary beam.

The primary beam plays an important role in the signal path from
sky sources to visibility data (or equivalently the synthesized image)
and the $l$-space filter function that determines the coupling of CMB
power to image variance depends on the antenna aperture illumination
function.  Therefore, instead of modelling the primary beam in the sky
plane, we have approximated the aperture illumination by a function
whose parameters characterize the central blockage and edge taper 
(James 1987) and have
used parameter values that give an antenna power pattern that fits
the measured pattern out to the second null.

The number of independent pixels in the sky image made with just the 
30.6-m baseline data, or equivalently the number of degrees of freedom in
the variance estimate, may be approximately the number of independent
visibility measurements along the 170~m long visibility track.  The first null
of the ATCA primary beam power pattern is at 7~arcmin radius at 8.7~GHz,
this implies that the aperture illumination has an autocorrelation function
with an effective diameter of about 17~m and hence the number of independent 
visibility measurements is about 10.  

\subsection{Telescope system noise}

The image variance resulting from the telescope thermal noise has been
measured by two methods.  First, by computing the rms over large sky 
regions well outside the primary beam area where the only contribution
to the image variance is expected to be the telescope thermal noise.  
Second, by separately imaging the visibilities
obtained as the XX and YY correlations between the two orthogonal linear
antenna signals X and Y and computing the rms over the (XX$-$YY)/2 image:
the Stokes I component of sources in the field cancel in this combined
image and if we assume that the Stokes Q, U and V flux densities are
negligible compared to the thermal noise, the system thermal noise will be
the only significant contributor to the variance in this combined image.
The two methods have yielded consistent values and the rms thermal noise
in the sky image is estimated to be 22.8~$\mu$Jy~beam$^{-1}$
in the image of the field made using the 30.6-m baseline data.
Adopting an antenna efficiency of 0.64 at 8.7~GHz (James 1985) 
the rms noise in the image is consistent with the system temperatures
being effectively 72 and 43.5~K respectively during the 1991 and 1994 epoch 
observations; the later value is in agreement with 1994 epoch measurements
of the system temperature made with microwave absorbers placed over the
feed horns (Gough 1994).

Along the visibility tracks in the $u,v$-domain, the signal-to-noise ratio
of the visibility data may vary because, for example, the 
total system temperature
may systematically vary with antenna elevation and hence with position
angle in the $u,v$-plane, or because data points may have 
different averaging times.  Visibility data are almost never weighted
proportional to the rms thermal noise; they are often weighted inversely
with the thermal noise variance.
Therefore, the image thermal
noise is described by a coherence function that is usually 
not derivable from
the synthesized beam.  To compute the angular coherence function of 
the thermal noise, or equivalently the distribution of thermal
noise power in the $u,v$-domain, we use images made well outside the primary
beam area:  we compute the 2D autocorrelation function of these sky images
and perform a 2D Fourier transform to the $u,v$-domain.  This provides us
with a $u,v$-domain filter appropriate for the telescope thermal noise.
Our simulations of the thermal noise component in the sky images have
assumed the noise to be Gaussian random, with a power spectrum identical to
this 2D $u,v$-plane filter function and normalized to give an image rms
noise same as the above-mentioned measured values.

The variance in the sky images have been determined by weighting the
pixel intensities with the primary beam attenuation. 
We have simulated images with solely the expected thermal noise contribution,
computed the weighted image variances and obtained their distribution
function. It may
be noted that because the variance
estimate is the weighted sum of squares of random variables, the variance
is not expected to follow a $\chi^{2}$ distribution.

\subsection{Foreground radio source confusion}

The radio sources in the sky are assumed to be Poisson distributed
and the differential source count $N(S)$ is expected to be
$$
	N(S) = 20.9 S^{-2.18}~{\rm arcmin^{-2}}~\mu{\rm Jy^{-1}}
		\eqno\stepeq
$$
\noindent at 8.7~GHz, where $S$ is the source flux 
density in $\mu$Jy (Partridge
et al. 1997 and references therein).  The ATCA Fourier-synthesis images 
have been
constructed without any zero-spacing visibility components and 
at every image pixel position the 
expectation for the flux density is zero despite the presence of
discrete sources in the sky.  The expectation of the 
variance contribution from these sources at any image position 
$(\theta_{\circ},\phi_{\circ})$ is given by
$$
	\sigma^{2} = \int\limits_{\theta,\phi} \vert b(\theta,\phi) 
                     \vert^{2} d \theta d \phi \int\limits_{y=y_{min}}
                     ^{y=y_{max}} y^{2} N(y) dy,
		\eqno\stepeq
$$
\noindent where $y_{min}$ and $y_{max}$ are the lower and upper limits to
the flux densities of sources present in the sky region and 
$b(\theta,\phi)$ represents a beam pattern that is the
product of the primary beam centred at the antenna pointing centre and
the synthesized beam centred at $(\theta_{\circ},\phi_{\circ})$.

We reckon, based on our 1.4-GHz survey of the region, that the field has
no sources exceeding about 0.5~mJy at 8.7~GHz.
The variance in the image made using the 30.6-m baseline data (Fig.~2) 
is a factor of two higher than that expected owing to these sources
assuming the Partridge et al. source counts.  However,
there is a reasonable (11 per cent) statistical probability that the `excess' 
variance in the field is foreground confusion.
The number of sources (3) detected with flux density 
exceeding 100~$\mu$Jy -- that constitute the model for confusion 
in the field -- and their combined flux density, 
are also a factor of two larger than expected.
A possible explanation for these is that our 3-component model
for confusion includes weaker sources that blend and give responses
above the detection limit.

Following the subtraction of the confusion model, the residual image
shown in Fig.~4 does not show evidence for any significant `excess'
variance that may be owing to residual weak foreground sources that have
not been included in the model.  For the assumed source counts, the 
expected variance contribution from sources weaker than 100~$\mu$Jy is
(50~$\mu$Jy beam$^{-1})^{2}$. The residual image
shown in Fig.~4 has a variance of (21.2~$\mu$Jy beam$^{-1})^{2}$ and  
the probability of observing a sample image with a variance as low as
this is $2 \times 10^{-4}$ if sources below 100~$\mu$Jy have remained
unsubtracted!  If we conjecture that the model derived from the 
61-m baseline visibility data has incorporated all sources above 25~$\mu$Jy,
corresponding to the rms thermal noise in the image made with these data,
we may then expect that the residual sources
in the field that are weaker than this limit may contribute a variance
of about (33~$\mu$Jy beam$^{-1})^{2}$ in images made from the residual 
30.6-m baseline data.  Even in this optimistic scenario, 
the probability of observing a sample with a residual variance as low
as that in the observed field is as small as 0.3 per cent!  

To summarize, the image variance appears to be in excess of the expectation
prior to any subtraction of sources, but after the confusion model is 
subtracted the image variance is below expectations that are derived
assuming a minimal residual and unsubtracted confusion. 
We are led to believe that either the derived confusion
model includes sources weaker than 25~$\mu$Jy,  
or perhaps that the faint sources are clustered.  
We are expanding our observations to several sky regions
and expect to have a better understanding of the contribution from
confusion following an examination of the residual fluctuations in
a larger sky area.  While deriving limits on CMB anisotropy in sections
5 and 6 based on our
observations of the field, we assume that the residual image 
has contributions from the telescope thermal noise and CMB anisotropy alone.

We have examined the possibility of using the skew coefficient $a_{3}$
as a possible indicator of residual foreground confusion. $a_{3}$ is defined as
$$ 
       a_{3} = {{m_{3}}\over{(\sqrt{m_{2}})^{3}}}, {\rm~with}
		~m_{3}=\sum x_{j}^{3},
                ~m_{2}=\sum x_{j}^{2}
	\eqno\stepeq
$$
\noindent and where the summations are over pixel 
intensities that are
weighted by the primary-beam attenuations at the pixel locations.  
We find that
the image in Fig.~2 has $a_{3}=0.13$ and the residual image in Fig.~4
has $a_{3}=-0.03$.  Our simulations of foreground confusion show that
the derived skew coefficient may be expected to have a standard deviation
of as much as 0.57 about the mean skew, and that the expectation for
the skew coefficient is just 0.11 and 0.05 respectively if foreground
sources below 100 and 25~$\mu$Jy remain unsubtracted.  Because of the
limited number of independent sky pixels in the image of the field, the
skew coefficient is not a useful statistic for characterizing the 
residual confusion; it may prove useful in statistical analyses of 
future observations that cover multiple fields. 

\subsection{CMB anisotropy}

We model the CMB anisotropy as a Gaussian random fluctuation in the
sky temperature that is completely described by its power spectrum
coefficients $C_{l}$.  The expectation for the image variance will then
depend on the telescope filter function (TFF) 
$F_{l}$ and the nature of the 
anisotropy power spectrum over the range in $l$-space that corresponds
to the pass-band of the filter. 

We assume that the anisotropy power spectrum is `flat' in $l(l+1)$-space
and that the $C_{l}$ coefficients are given by
$$
	C_{l} = (Q_{flat})^{2} \left({{24 \pi}\over{5}} 
		{1\over{l(l+1)}} \right),
	\eqno\stepeq
$$
\noindent where $Q_{flat}$ denotes the quadrupole normalization of the
power spectrum.  The temperature variance in the sky image
is then expected to be
$$
	(\Delta T)^2 = \sum_{l} {{(2l+1)C_{l}}\over{4 \pi}} F_{l}
	\eqno\stepeq
$$
\noindent and this leads to the expression in equation~(1).  

We compute
the image variance as a weighted mean of the squares of the pixel 
intensities (image pixel intensities are in units of Jy~beam$^{-1}$)
using weights for the intensities that are the primary-beam
attenuations at the pixel positions.  As discussed in Appendix~A, we have
computed the ATCA TFFs
at the different image pixel locations using equation~(A4) 
and averaged them, using weights that are again the primary-beam
attenuations, to obtain the filter function $F_{l}$, in units of 
(Jy~beam$^{-1}$~K$^{-1})^{2}$, appropriate to the computed image variance.
We have computed the $F_{l}$'s using the ATCA
synthesized beams corresponding separately to the images made with the 
30.6 and 61-m baseline data, they are shown in Fig.~6.
In the large-$l$ regime of the ATCA $F_{l}$, 
the expected image variance (in units of (Jy~beam$^{-1})^{2}$)
is approximately given by
$$
	\Delta S^{2} = (Q_{flat})^{2} \sum_{l} {{12}\over{5l}} F_{l}.
	\eqno\stepeq
$$

\beginfigure*{6}
  \nofloat
  \epsfbox[30 70 430 330]{fig6.ps}
  \caption{{\bf Figure 6.} ATCA telescope filter function.
   The continuous
   line shows the TFF corresponding to the image made with the 30.6-m
   baseline data, the dashed line shows the 
   filter corresponding to the 61-m
   baseline data.}
\endfigure

The TFF 
corresponding to the 30.6-m baseline data, shown as the continuous line
in Fig.~6,
is the filter function corresponding to the residual image shown in
Fig.~4.  It has a peak of 0.48 at $l=4700$ and has half-maximum values
at $l=3350$ and 6050.  Summation over $l$-space yields
$$
	\sum_{l} {{12}\over{5l}} F_{l} = 0.765~
        \left( { {\mu{\rm Jy~beam}^{-1}} \over {\mu{\rm K}} }\right)^{2}.
	\eqno\stepeq
$$
\noindent If we adopt a `flat' CMB anisotropy spectrum with normalization
$Q_{flat}=18~\mu$K corresponding to the detected power at multipoles 
$l \la 20$ (4-yr {\it COBE DMR} results in Bennett et al. 1996), the expected 
variance contribution in the 30.6-m baseline image is expected to be
$18^{2} \times 0.765 = (15.7~\mu$Jy~beam$^{-1})^{2}$.  

A Fourier synthesis telescope with finite-aperture elements
measures visibilities that may be considered to be   
the convolution of the all-sky visibility with the autocorrelation
of the antenna aperture illumination.  
The visibility measurements may be viewed as samples 
along 1D $(u,v)$-tracks of this 2D ($u,v$)-plane visibility function.
Because the ATCA antenna aperture
extends 11~m in radius, its autocorrelation function will
extend to a radius of 22~m.  It may be noted that this autocorrelation
function has a diameter of 44~m which exceeds the spacing between the
1D $(u,v)$-tracks corresponding to the 30.6 and 61~m baselines.
Therefore, the visibility data along the 30.6 and 61-m
baseline visibility tracks are individually averages over 
regions of the all-sky visibility function and these regions mutually
overlap.
This results in a partial overlap between the TFFs corresponding
to the 30.6 and 61-m visibility data.
At any instant, there are three 61-m baselines as
compared to four 30.6-m baselines; therefore, the 61-m baseline image is 
less sensitive to unresolved foreground sources
as compared to the 30.6-m baseline data, but by only a factor 0.87.
The peak of the TFF corresponding
to the 61-m baseline data is 0.087, a factor 0.18 of the peak of the 30.6-m
baseline TFF.  Summing over the 61-m baseline TFF, 
$\sum 12F_{l}/(5l) = 0.085$: CMB anisotropy with a flat spectrum will
be expected to contribute a variance in the 
61-m baseline image that is a factor
0.11 of the variance in the 30.6-m baseline image.  Adopting a
normalization $Q_{flat}=18~\mu$K, the 61-m baseline image is expected to
have a variance contribution $(5.2~\mu$Jy~beam$^{-1})^{2}$ from CMB
anisotropy.   Because (a) the two TFFs have only a small overlap and
are largely sampling CMB power at different multipole ranges and
(b) the 61-m baseline TFF has a greatly reduced sensitivity to
CMB anisotropy, the confusion
model that is derived from the 61-m baseline data is unlikely to
significantly reduce the CMB anisotropy power expected in the
residual 30.6-m baseline image.

We have simulated sky images by generating `flat' spectrum CMB
power in the $(u,v)$-plane: pixels of size $\delta u,~\delta v$ that
are at distance $\sqrt{u^2 + v^2}$ from the centre of the $(u,v)$-plane
are given Gaussian random complex conjugate visibilities with variance 
$$
	(\Delta T)^{2} = (Q_{flat})^{2} \left
	     ({6 \over {5 \pi l^{2}}} \right) \delta u \delta v.
	\eqno\stepeq
$$
\noindent The $(u,v)$-plane CMB anisotropy model is then filtered
by the TFF corresponding to the 30.6-m baseline data and inverted
to form simulated sky images.  We have thereby obtained distributions
of the sample variance from simulated images that have variance contribution
from CMB anisotropy alone.  It may be noted here that the radial distribution
of power in the $(u,v)$-plane 
in the case of our `flat' model CMB anisotropy differs from
that for thermal noise and, therefore, the two contributions differ
in the number of degrees of freedom in their contributions to image variance.

\section{Limits on CMB flat-band power}

\beginfigure*{7}
  \nofloat
  \epsfbox[75 95 580 400]{fig7.ps}
  \caption{{\bf Figure 7.} Likelihood functions. The dashed line corresponds
	to the null hypothesis $\bbbh_{1}$, the continuous curve corresponds
        to the hypothesis $\bbbh_{2}$ that $Q_{flat}=22.4~\mu$K.  The dotted
        line corresponds to the distribution function assuming that 
        `flat' spectrum CMB anisotropy with $Q_{flat}=22.4~\mu$K is the only
        contributor to the image variance.}
\endfigure

We use the likelihood-ratio test to derive limits on possible
CMB flat-band power in the data.  We assume that the residual image
has contributions only from the telescope thermal noise and possibly
CMB anisotropy.  The observed variance in the
residual image shown in Fig.~4 is 
$$
	\sigma_{obs}^{2} = (21.2~\mu$Jy~beam$^{-1})^{2}.
	\eqno\stepeq
$$

We adopt the null hypothesis $\bbbh_{1}$ that the variance
contribution from CMB anisotropy is zero.
$$
	\bbbh_{1}~:~\sigma_{CMB}^{2} = 0.
	\eqno\stepeq
$$
Simulations of sky images that have only thermal noise with
an expected variance
$(22.8~\mu$Jy~beam$^{-1})^{2}$ then yields the probability distribution
for the sample variance.  We show this distribution as the dashed line 
in Fig.~7 and represents the likelihood function 
$\bbbp(\sigma_{obs}^{2} \vert 0)$ which is the probability of obtaining
any observed variance conditional on $\bbbh_{1}$.

We next hypothesize that the sky has flat-band CMB anisotropy 
quantified by the normalization $Q_{flat}$, {\it i.e.,}
$$
	\bbbh_{2}~:~\sigma_{CMB}^{2}~{\rm corresponding~to~} Q_{flat}.
	\eqno\stepeq
$$
We simulate sky images with purely
CMB anisotropy to obtain the distribution function for its 
variance contribution;  we show
this distribution in Fig.~7 as a dotted line for the specific
choice $Q_{flat}=22.4~\mu$K.  
The distribution function for the observed image variance, with
contributions from thermal noise and CMB anisotropy,  will be a 
convolution of the two individual probability distributions.
This distribution, shown as a continuous curve in Fig.~7,  
represents the likelihood
function $\bbbp(\sigma_{obs}^{2} \vert Q_{flat})$ which is the probability
of obtaining any observed variance conditional on $\bbbh_{2}$.

The size $\alpha$ of the test 
is the probability of rejecting $\bbbh_{2}$ when it is true (a type {\sc i}
error).  For the choice $Q_{flat}=22.4~\mu$K, $\alpha=0.05$ and, 
therefore, $Q_{flat}=22.4~\mu$K is a 95 per cent confidence upper limit.

The power $\beta$ of the test is defined as: $\beta$ = 1 $-$ probability
of accepting $\bbbh_{2}$ when $\bbbh_{1}$ is true (a type {\sc ii} error).
For our choice of $Q_{flat}=22.4~\mu$K, the power $\beta = 0.47$.  
In order to increase the power of the test to a value $\beta = 0.5$,
we may increase the decision variance to 
$\sigma_{obs}^{2} = (21.8~\mu$Jy~beam$^{-1})^{2}$ and change the
hypothesis $\bbbh_{2}$ to correspond to a choice 
$Q_{flat}=23.6~\mu$K.  For this choice of parameters, 
the test will reject $\bbbh_{2}$ with
size $\alpha = 0.05$ (95 per cent confidence) and the test will
simultaneously have a power $\beta = 0.5$.  

To summarize, the ATCA observations of the field place an upper limit
of $Q_{flat} < 23.6~\mu$K with 95 per cent confidence in an 
$l$-space filter that peaks at $l=4700$.  The filter has half-maximum
values at $l = 3350$ and 6050.

\section{Limits on CMB anisotropy with a Gaussian-form autocorrelation function}

\begintable*{1}
\caption{{\bf Table 1.} ATCA limits on GACF anisotropy models.}
\halign{ \hfil#\hfil & \quad\hfil#\hfil\quad & \quad\hfil#\hfil\quad 
& \quad\hfil#\hfil\quad & \hfil#\hfil \cr
Coherence scale  & Size of the test & Power of the test & CMB temperature rms & 
$\Delta T/T$ \cr
$\xi_{c}$ & $\alpha$ & $\beta$ & $C_{\circ}^{1\over2}$ & 
$C_{\circ}^{1\over2}/T_{\circ}$ \cr
0\farcm5 & 0.05 & 0.44 & $58~\mu$K & $2.1 \times 10^{-5}$ \cr
0\farcm5 & 0.05 & 0.50 & $63~\mu$K & $2.3 \times 10^{-5}$ \cr
1\farcm0 & 0.05 & 0.44 & $42~\mu$K & $1.5 \times 10^{-5}$ \cr
1\farcm0 & 0.05 & 0.50 & $45~\mu$K & $1.6 \times 10^{-5}$ \cr
2\farcm0 & 0.05 & 0.44 & $63~\mu$K & $2.3 \times 10^{-5}$ \cr
2\farcm0 & 0.05 & 0.50 & $68~\mu$K & $2.5 \times 10^{-5}$ \cr
}
\tabletext{$T_{\circ} = 2.73$~K}
\endtable

The CMB anisotropy is sometimes assumed to have a Gaussian
autocorrelation function (GACF).  The sky temperature is modelled to have
a Gaussian distribution with zero mean and the autocorrelation
 of the sky temperature
is taken to be of the form
$$
	C(\xi) = C_{\circ}~e^{-\left({{\xi^{2}}\over{2\xi^{2}_{c}}}\right)}.
		\eqno\stepeq
$$
Parameter $C_{\circ} = \langle T^{2}(\theta,\phi) \rangle$ represents the
variance of the sky temperature and parameter 
$\xi_{c} = \left(-C(0)/C^{\prime\prime}(0)\right)^{1\over2}$ 
represents the coherence scale ($C^{\prime\prime}(\xi)$ denotes the second
derivative of the autocorrelation function).  
Assuming that the coherence scale is a small
angle, the power spectrum of the
CMB temperature anisotropy may be approximately represented as $l$-space
coefficients
$$
	C_{l} = 2 \pi C_{\circ} \xi^{2}_{c} e^{- \left( {{l^{2} \xi^{2}_{c}} \over 
	         {2}} \right)}.
	\eqno\stepeq
$$
This model may be expected to contribute an image variance 
$$
	(\Delta S)^{2} = {1 \over 2} C_{\circ} \xi^2_c \sum_{l} (2l+1) e^{
	                  -\left({ {l^{2} \xi^{2}_{c}}\over{2} }\right)} F_{l}
	\eqno\stepeq
$$
to the ATCA residual sky image.  In this expression, $\xi_{c}$ is the coherence
scale in radians and $F_{l}$ is the ATCA TFF 
corresponding to the variance estimate
in the residual image of the observed field.

Assuming that the sky temperature variance $C_{\circ}$ is an invariant, the
expected image variance $(\Delta S)^{2}$ is a 
maximum for an anisotropy model with $\xi_{c} = 1$~arcmin.  
We have simulated sky images with the CMB anisotropy 
modelled to have a GACF
and derived the distribution function for the image variance contribution.
Limits to the model parameter $C_{\circ}$ may be derived for specific choices
of the coherence scale $\xi_{c}$ using the likelihood ratio test described
in the previous section. 

Adopting a value $\xi_{c} = 1$~arcmin,  we find that the observed image variance
of $\sigma_{obs}^{2} = (21.2~\mu$Jy~beam$^{-1})^{2}$ implies a 95 per cent confidence
upper limit of 42~$\mu$K on $C_{\circ}^{1\over2}$.  This result corresponds to a 
test with size $\alpha = 0.05$ and power $\beta = 0.44$.  Requiring that the test
have a power $\beta$ of 
at least 0.5 relaxes the 95 per cent confidence upper limit
to a value $C_{\circ}^{1\over2} = 45~\mu$K.  We have tabulated these limits along
with those for models with coherence scales of 0.5 and 2~arcmin in Table~1.

\section{A comparison with other measurements of CMB anisotropy}

The best upper bounds
reported to date in observations with the 
Owens Valley radio observatory (OVRO) is that the fractional
fluctuations in the CMB sky is limited to 
$\Delta T/T < 1.7 \times 10^{-5}$ at a resolution of 2~arcmin;
$C_{\circ}^{1\over2}/T_{\circ} < 1.9 \times 10^{-5}$ in the case of
fluctuations with a GACF and a coherence scale of 2\farcm6 (OVRO
NCP experiment: Readhead et al. 1989).  
Sensitive imaging observations with the Very Large Array (VLA) at 
8.4~GHz have been used to place a limit of $2.0 \times 10^{-5}$ on 
the fractional temperature fluctuations ($\Delta T/T$) 
in the CMB at a resolution
of 1~arcmin (VLA experiment: Partridge et al. 1997). 
Assuming that the CMB anisotropy has a GACF form, an upper limit of 
$\Delta T/T \lid 2.1 \times 10^{-5}$ has been set for a coherence scale
of 1\farcm1 based on observations at 142~GHz using a 6-element bolometer 
array (SuZIE experiment: Church et al. 1997).   The SuZIE observations
have also been
used to derive a 2-$\sigma$ upper limit of $Q_{flat} < 26~\mu$K 
at an effective $l \sim 2340$ (Ganga et al. 1997). 
Among these three observations, the first two have sensitivities per
resolution element that are comparable to our ATCA observations.
The SuZIE observations have covered a larger sky region.  
The ATCA observations presented in this work
limit $Q_{flat}$ to 
$23.6~\mu$K at $l_{eff} = 4700$ and limit $C_{\circ}^{1\over2}/T_{\circ}$
to $1.6 \times 10^{-5}$ for GACF form CMB anisotropy with $\xi_{c} = 1$~arcmin.
These different
experiments attempting to measure the CMB anisotropy on arcmin scales
have comparable upper limits on random phase CMB anisotropy at
multipoles $l > 1000$.

\section*{Acknowledgments}

The Australia Telescope is funded by the Commonwealth of Australia for
operation as a National Facility managed by CSIRO.
RS thanks Rajaram Nityananda for his interest and helpful suggestions.

\appendix
\section{Telescope filter functions (TFF)}

In the case of CMB anisotropy experiments that are done using single-dish 
telescopes with a specified beam switching scheme, the telescope beam
is usually described by a 2D function on the sky that is normalized to
unit volume.  The antenna temperature measured at any sky position is then
the mean sky brightness temperature weighted by the beam pattern.  The 
antenna temperatures obtained at a set of sky positions (that are
defined by the beam switching scheme) are combined linearly to form
an estimate of the CMB anisotropy. The `effective' beam on the sky 
is the same linear combination of the 2D telescope beam patterns.
The TFF corresponding to the estimate of the 
CMB anisotropy that is made with the `effective' beam is
the spherical harmonic decomposition of the
`effective' beam pattern.

In observations for CMB anisotropy made with Fourier synthesis
telescopes, the sky temperature anisotropy is viewed
by the array through the element primary beam pattern.  The
temperature anisotropy, attenuated by the primary beam pattern of
the individual array element, is convolved by the synthesized beam
pattern.  The synthesized beam is normalized to peak unity and has
zero volume.  Unlike the case of the `effective' beam in beam-switched,
single-dish observations, the synthesized beam is not usually decomposable
into a linear combination of identical (position shifted) beams that have
finite volume.  Therefore, the measurements (image pixel intensities)
in Fourier synthesis images are integrals of the sky brightness
temperature over the synthesized beam and cannot be converted to
mean (or weighted mean) temperatures by normalising with any beam
volume.  The measurements will be in units of flux density (Jy)
per beam rather than temperature.

For large multipole orders and small angles, the spherical harmonic
decomposition may be approximated by a continuous Fourier transform
and the $C_{l}$'s may be related to the sky autocorrelation function
$C(\theta)$ by
$$
	C_{l} = 2 \pi \int\limits_{0}^{\infty} \theta d \theta C(\theta)
                J_{\circ}(l \theta).
        \eqno\stepeq
$$
\noindent Multipole order $l/(2 \pi)$ and angular distance $\theta$ 
(in radians) are Fourier transform conjugates.  If the `effective' beam
is a purely radial function $b(\theta)$, the TFF
will be similarly given by
$$
	F_{l} = \left| 2 \pi \int\limits_{0}^{\infty} \theta d \theta 
                b(\theta) J_{\circ}(l \theta) \right|^{2}.
        \eqno\stepeq
$$
\noindent If the `effective' beam is not simply a radial function, one
computes the 2D Fourier transform $\bbbf(l,\xi)$ of the 2D beam 
$b(\theta,\phi)$ and circumferentially averages 
$\vert \bbbf(l,\xi) \vert ^{2}$ to get the filter function
$F_{l}$.   

The TFF in Fourier synthesis images will vary
across the image.  At any image pixel location 
$(\theta_{\circ},\phi_{\circ})$, we first determine
the `effective' beam $b(\theta,\phi)$ as the product of the primary beam
centred at the antenna pointing centre and the synthesized beam centred
at  $(\theta_{\circ},\phi_{\circ})$. The two beams are separately
normalized to have peak unity.  The beam is Fourier transformed to give
$$
  	\bbbf(l,\xi) = \int\int b(\theta,\phi) e^{i(\theta l
        cos\xi + \phi l sin \xi)} d\theta d\phi.
        \eqno\stepeq
$$
\noindent The TFF is then derived as
$$
	F_{l} = {1 \over {2 \pi}} \int\limits_{0}^{2 \pi} 
		\vert \bbbf(l,\xi) \vert^{2} d\xi 
		\left({{2k}\over{\lambda^2}}\right)^{2}.
	\eqno\stepeq
$$
\noindent $F_{l}$, as defined in this form for Fourier-synthesis
imaging,  has units (Jy~beam$^{-1}$~K$^{-1})^{2}$.  

If the image variance is computed as a linear combination of the squares of 
the pixel intensities (a weighted sum of squares), the TFF for the variance
estimator may correspondingly be computed as the same linear combination
of the $F_{l}$'s  evaluated at the individual pixel locations. 

\end